\documentclass[prd,twocolumn,superscriptaddress,floatfix,amsmath,amssymb,amsfonts,longbibliography,nofootinbib]{revtex4-2}

\usepackage{float}

\usepackage{comment}
\usepackage[normalem]{ulem}
\usepackage[english]{babel}
\usepackage{graphicx}
\usepackage{dcolumn}
\usepackage{bm}
\usepackage{blindtext}
\usepackage{verbatim}
\usepackage{mathrsfs}
\usepackage{musicography}
\usepackage{amsmath}
\usepackage{dsfont}
\usepackage{mathrsfs}
\usepackage{amssymb}
\usepackage{amsthm}
\usepackage{cancel}
\usepackage{physics}
\usepackage{epstopdf}
\usepackage{mathtools}
\usepackage{color}
\usepackage[usenames,dvipsnames]{pstricks}
\usepackage{epsfig}
\usepackage{pst-grad} 
\usepackage{pst-plot} 
\usepackage{hyperref}
\usepackage{verbatim}
\usepackage{afterpage}
\usepackage{sidecap}
\hypersetup{
    colorlinks=true,
    linkcolor=blue,
    filecolor=red,      
    urlcolor=cyan,
}
\usepackage{tikzsymbols}

\newcommand{\kittyket}
{\ket{\raisebox{-.43ex}{\SchrodingersCat{1}}\!\!}}

\begin{document}
\title{\textbf{Magnetically Induced Schr\"{o}dinger Cat States: The Shadow of a Quantum Space}}
\author{Partha Nandi$^a$}
\email{pnandi@sun.ac.za}
\author{Nandita Debnath$^b$~$^e$}
\email{debnathnandita14@gmail.com }
\author{Subhajit Kala$^c$}
\email{s.kala@iitg.ac.in}
\author{A. S. Majumdar$^d$}
\email{archan@bose.res.in}
\affiliation{$^a$Department of Physics, University of Stellenbosch, Stellenbosch-7600, South Africa.\\
$^b$Department of Physics, University of Calcutta, Kolkata 700009, India.\\
$^e$School of Physical Sciences, Indian Association for the Cultivation of Science, Kolkata 700032, India.\\
$^c$Department of Physics, Indian Institute of Technology Guwahati, Guwahati 781039, Assam, India.\\
$^d$S. N. Bose National Centre for Basic Sciences,
JD Block, Sector III, Salt Lake, Kolkata 700106, India.}

\begin{abstract}
Schr\"{o}dinger cat states, which are superpositions of macroscopically distinct states, are potentially critical resources for upcoming quantum information technologies. In this paper, we introduce a scheme to generate entangled Schr\"{o}dinger cat states in a non-relativistic electric dipole system situated on a two-dimensional plane, along with an external potential and a uniform strong magnetic field perpendicular to the plane. Additionally, our findings demonstrate that this setup can lead to the phenomenon of collapse and revival of entanglement for a specific range of our model parameters. 
\end{abstract}
\maketitle

\section{Introduction}

In quantum  theory, the transition between the microscopic and macroscopic worlds is one of the less-understood features \cite{Zuk}. Such a transition
plays a direct role  in the realm of quantum measurements.  In an ideal measurement paradigm, the interaction of macroscopic equipment and a microscopic system yields entanglement and a superposed quantum  state with both macroscopic and microscopic components \cite{pn2}. Schr\"{o}dinger was the first to highlight the
physical subtleties of this kind of superposition by replacing the macroscopic part of the system by a ``cat", in order to illustrate a dramatic superposition of ``states" of both alive and dead cats, that should, in
practice,  be distinguished macroscopically \cite{par}.
The superposition of macroscopically different quantum states, 
generically referred to as non-classical Schr\"{o}dinger Cat State (SCS) \cite{par,2,bose}, is crucial for understanding the conceptual underpinnings  of quantum physics, especially with reference to wave function collapse models \cite{pnr,pr,hr,hr1}. In recent years, the advancement of quantum technologies    
has brought into sharp focus the utility of several quantum phenomena such as photon anti-bunching \cite{Rov}, sub-Poissonian statistics \cite{mar} and squeezing \cite{mtw}, along with the dynamics of SCS.

The success of quantum information theory and its potential applications 
\cite{qinform, qinf} that significantly outperform their classical equivalents have recently sparked a renewed interest in the generation of non-classical states such as SCS.  
Several applications of cat states have been suggested in the realm of  
quantum information \cite{qinfor}, 
 quantum metrology \cite{qmeter}, quantum teleportation \cite{qteleport},  and quantum error correction schemes \cite{qerror1, qerror2}. Besides, the concept of decoherence between two superposed quantum objects, or the quantum-to-classical transition, can  be studied using the SCS as a platform. In quantum optics, a superposition of two diametrically opposite coherent states $\ket{\pm\alpha}$ with large value of $\mid\alpha\mid$, can be interpreted as a quantum superposition of two macroscopically distinct states, {\it i.e.}, a Schr\"{o}dinger cat-like state \cite{tw,legg}. However,  due to decay of their interference properties,  it is extremely difficult to detect such states in 
practice \cite{egg}. Nonetheless, the universality of SCS enables it to
be realized in a wide variety of physical arenas such as nonlinear quantum optics \cite{Sb}, quantum dot systems  \cite{kaku}, superconducting cavities \cite{kau}, Bose Einstein condensates (BEC) \cite{goenner} and  quantization of weak gravity \cite{rbm,wolf,sch}. A fascinating direction of research in recent years has been the mechanism for the natural generation of SCS in some specific condensed matter systems \cite{sch1,sch2}.
 
 Schr\"{o}dinger cat states with entanglement based protocols provide a novel technique to explore short-distance quantum physics in a non-relativistic domain when there is a magnetic dipole interaction background \cite{ed}. At extremely short distances, the space-time structure needs to be ``granular"  in order to account for both  gravity and quantum uncertainty \cite{ein}.  A
 viable approach towards quantum gravity is through quantizing space-time itself \cite{in}, rather than the construction of an effective field theory of
 gravity. This approach is an active area of  research on quantum gravity, commonly referred to as non-commutative geometry \cite{go,pe}. The fundamental goal is to derive classical geometry from a suitable limit of a non-commutative algebra. Though such a proposal may appear as ad-hoc \cite{pek},  the physical justification for such a non-commutative space-time is strong since it provides a solution to the geometric measurement problem  near the Planck scale. 
 
 Non-commutative geometry appears naturally in various non-relativistic planar systems. For instance, it occurs using the lowest Landau-level (LLL) projection to study the behavior of charged particles in a strong magnetic field \cite{ek}. Further, the incompressibility of fractional quantum Hall fluids \cite{qhefluids} has a strong connection to the emergence of a non-commutative geometry in which the fundamental Planck length is substituted by the magnetic length. Non-commutative space-time forms  an alternative paradigm for studying the behavior of relativistic anyonic systems in interaction with the ambient electromagnetic field \cite{vpn, Rabin}. Additionally,  non-commutative properties of real-space coordinates in the presence of the Berry curvature \cite{k} produce  skew scattering by a non-magnetic impurity without relativistic spin-orbit interactions in a condensed matter system. Non-commutative space provides a paradigm for describing the behavior of the quantum to classical  transition under the influence of decoherence \cite{fg1,fg2}, which is relevant for implementation of quantum information
protocols.  From an experimental standpoint,  there have been efforts in search of evidence of possible non-commutative effect manifestations in cosmology and high-energy 
physics \cite{kk,c,d}. A testable framework has been suggested in low-energy experiments  in the arena of quantum Hall effect \cite{qhe,e}. 

The motivation for the present study is to investigate whether multi-component entangled non-classical SCS could be produced in  deformed quantum space,
where non-commutativity arises naturally in an easily accessibly low energy
physical system.  In this article, we investigate the phenomenology of a two-particle electric dipole model with an additional harmonic interaction and a strong background magnetic field, with motion constrained to the plane perpendicular to the field. Such a system may be considered as a toy  version of a real Excitonic dipole set-up \cite{exci}. By exploring the high magnetic field limit, we reveal the emergence of planar  non-commutative space as a natural consequence. Furthermore, we establish the deformed Heisenberg algebra as the origin of multi-component entangled SCS in this system. Moreover, we quantify the degree of entanglement of our SCS, and show that the phemomenon of collapse and revival of entanglement \cite{R1, yueberly, R2} occurs in this system under the influence of the harmonic
potential.  

The organization of our paper is as follows. The interacting two particle electric dipole system is introduced in Section 2,  showing how classical non-commutative  space appears in the presence of a very strong, constant, uniform magnetic field. Then, in Section 3, we move on to the quantum picture, where
intricacies of the system dynamics are revealed, in context of mapping
between two reference frames.  Section 4 discusses how our model with a harmonic oscillator potential that is dependent only on one spatial variable is able to generate entangled multi-component Schr\"{o}dinger cat states. In Section 5, we compute the degree of entanglement in the generated SCS system and
demonstrate  that it exhibits the phenomenon of entanglement collapse and revival.  Section 6 is reserved for concluding remarks and discussions.  There are three appendices. The first appendix discusses the phase space algebra in the center-of-mass and relative coordinate frames for unequal test masses. The aim of the second appendix is to derive the explicit structure of the purity function. The third appendix is reserved for a generalization of our system’s potential and its consequences.

\section{ Two-Particle System:  Classical picture}

We begin by considering a pair of non-relativistic, oppositely charged particles with equal mass $m$ moving on the plane subjected to a constant magnetic field $B$ along the $z$ axis (ignoring Coulomb and radiation effects). In component form, $x_i$ and $y_i$ $(i = 1, 2)$ correspondingly represent the positive and negative charge coordinates. The coordinate $z$ can be suppressed because the dynamics of the system are confined to a plane.

\begin{figure}[h]
    \centering
    \includegraphics[scale=0.9]{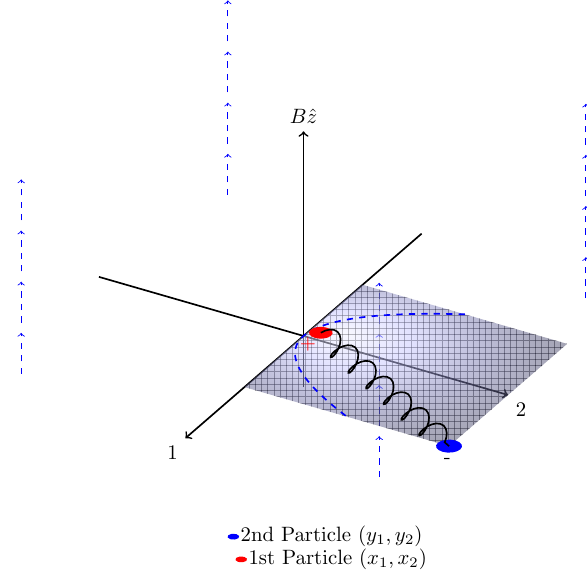}
    \caption{This figure shows a magnetic field (dashed blue lines) aligned along the \( z \)-axis, represented by field lines at different \( z \)-coordinates. The shaded parabolic potential well along the \( 1 \)-axis is depicted with a grid pattern, indicating its shape and position. A positively charged particle (red circle, labeled “+") is located within the well, while a negatively charged particle (blue circle, labeled “-") is outside the well. These particles are connected by a spring (black wavy line), symbolizing an interaction between them. The dashed blue curve within the shaded region represents the potential profile in the \( x_{1} \)-direction.}
    \label{fig:enter-label}
\end{figure}

Standard Lagrangian in C.G.S. units is used to define the system as follows \cite{Dunne,bag,pn}:
\begin{align}
L=\frac{1}{2}m(\dot{x}_{i}^{2}+\dot{y}_{i}^{2})+\frac{eB}{2c}\epsilon_{ij}(x_{j}\dot{x}_{i}-y_{j}\dot{y}_{i})\notag\\
-\frac{K_{0}}{2}(x_{i}-y_{i})^{2}-V({x_{1}});~~~i,j=1,2
\label{l}
\end{align}
where $c$ is the speed of light in a vacuum, $\epsilon_{ij}$ is the Levi-Cevita symbol, and $K_0$ is the spring constant corresponding to the harmonic interaction between the two oppositely charged particles. This model  is constructed in the spirit of the ``2D excitonic dipole model" \cite{exmol,emg,dipex}, wherein $m$ can be realized by the effective mass
of the ``electron-hole" pair in some specific cases where the magnitude of the effective mass of electrons and holes can be considered as approximately same and the Fermi velocity provides an upper bound for its characteristic velocity in a real physical solid state system. Note that the first term of the above Lagrangian (\ref{l}) represents the kinetic term of the charges and the second term represents their interaction with the external magnetic field $\vec{B}$. We use a rotationally symmetric gauge to define the vector potential $\vec{A}$ satisfying the equation $\vec{\nabla}\times\vec{A}=B\hat{z}$. The third term is the harmonic interaction between the two charges, and finally, the fourth term describes the additional interaction of the positive charge with an impurity in the $x_{1}$ direction. The limit of a strong magnetic field $B$ and small mass $m$ such as $\frac{m}{eB}\rightarrow 0$ is of interest here, in which  the kinetic energy term becomes negligible in the Lagrangian (\ref{l}) \cite{BD}. Thus, we may approximate  the dynamics by the effective Lagrangian:
\begin{equation}
L_{eff}=\frac{eB}{2c}\epsilon_{ij}(x_{j}\dot{x}_{i}-y_{j}\dot{y}_{i})-V_{0}(x_i,y_i),
\label{e1}
\end{equation}
where $V_{0}(x_i,y_i)=\frac{K_{0}}{2}(x_{i}-y_{i})^{2}+V({x_{1}})$.

The Lagrangian equations of motion of the co-ordinates of the positive and negatively charged particles are given by,
\begin{equation}
\dot{x}_{i}=\frac{c}{eB}\epsilon_{ij}\frac{\partial V_{0}}{\partial x_{j}},~~ \dot{y}_{i}=-\frac{c}{eB}\epsilon_{ij}\frac{\partial V_{0}}{\partial y_{j}}
\end{equation}
Since our effective Lagrangian (\ref{e1}) is  in first-order form, the  effective Hamiltonian of the model is given by
\begin{equation}
H_{eff}=V_{0}(x_{i},y_{i})
\label{h}
\end{equation}
In order to show the equivalence between the Lagrangian and Hamiltonian formalism \cite{fj,Rb}, we consider Hamilton's equations of motion:
\begin{equation}
\dot{x}_{i}=\{{x_{i},H_{eff}}\}_{SB}={\{x_{i},V_{0}(x_{i},y_{i})}\}_{SB}
\label{a1}
\end{equation}
\begin{equation}
\dot{y}_{i}=\{{y_{i},H_{eff}}\}_{SB}={\{y_{i},V_{0}(x_{i},y_{i})}\}_{SB}
\label{a2}
\end{equation}
where $\{,\}_{SB}$ denotes the classical symplectic brackets.
The nontrivial symplectic structure can be readily obtained by comparing the Lagrangian equations of motion (3) with the form of Hamilton's equations of motion ($\ref{a1},\ref{a2}$), yielding the following brackets:
\begin{equation}
\{x_{i},x_{j}\}_{SB}=\frac{c}{eB} \epsilon_{ij};~~\{y_{i},y_{j}\}_{SB}=-\frac{c}{eB} \epsilon_{ij};~~\{y_{i},x_{j}\}_{SB}=0
\label{c}
\end{equation}
The canonical spatial translation generators for individual charged particles are given by
\begin{equation}
P_{x_{i}}=\frac{eB}{c}\epsilon_{ij}x_{j};~~P_{y_{i}}=-\frac{eB}{c}\epsilon_{ij}y_{j}
\end{equation}
Using the above expressions and the nontrivial symplectic structures between the position co-ordinates ($\ref{c}$), it can  be checked that the momentum co-ordinates also satisfy a nontrivial symplectic bracket, given by 
\begin{align}\label{sym1}
\{P_{x_{i}},P_{x_{j}}\}_{SB}=\frac{eB}{c} \epsilon_{ij};~~\{P_{y_{i}},P_{y_{j}}\}_{SB}=-\frac{eB}{c}\epsilon_{ij};~~\notag\\ \{x_{i},P_{x_{j}}\}_{SB}=\{y_{i},P_{y_{j}}\}_{SB}=\delta_{ij}
\end{align}

In our model, it may be noted that we have considered a sufficiently strong magnetic field such that the kinetic energy of the charged particles is not dominant compared to the magnetic term. 
In the ultra-strong magnetic regime, the kinetic energy becomes negligible, and its effects are entirely suppressed \cite{er,r}. 
The energy scale associated with a strong magnetic field corresponds to the shortest distance, and one may think that the motion of the charged particle at this length scale may be treated relativistically. However,  the magnetic quantum length scale is $
l_B = \sqrt{\frac{\hbar c}{eB}} \approx 10^{-3} \text{m}$
where the magnetic field \( B \) is on the order of \( \sim 10^3 \text{ Gauss} \). The Compton wavelength is given by 
$
\lambda_c = \frac{h}{m_e^* c}$
where \( m_e^* \) is the effective mass of the electron, ranging from 0.01 to 10 times the mass of a free electron \( (m_e) \). 
The maximum Compton wavelength \( (\lambda_{c})_{max} \) is approximately \( 10^{-10} \text{ m} \), assuming \( m_e^* = 0.01 m_e \) \cite{r+}, which is much smaller compared to \( l_B \). It is well known that relativistic effects become significant when the length scale associated with the particle is comparable to or in the vicinity of the Compton wavelength of the particle (see \cite{cr} for further details). 
Therefore, the relativistic nature of our toy model can be completely neglected, and it can be treated as a non-relativistic model.

\section{Quantum dynamics}

In this section, we discuss the quantum theory of our non-relativistic two-particle model at the strong magnetic field limit  by elevating the phase space variables to the level of quantum operators. We  obtain the nontrivial or unusual commutation brackets 
$([,]=i\hbar\{,\}_{SB})$ between the position operators given by:
\begin{equation}
[\hat{x}_{i},\hat{x}_{j}]=i l^{2}_{B}\epsilon_{ij};~~[\hat{y}_{i},\hat{y}_{j}]=-il^{2}_{B}\epsilon_{ij};~~[\hat{x}_{i},\hat{y}_{j}]=0;~~i,j=1,2
\end{equation}
with $l_{B}=\sqrt{\frac{\hbar c}{eB}}$  known as the magnetic  quantum length scale. 
Likewise, the  other nontrivial phase space non-commutative algebras are given as 
\begin{equation}\label{sym1}
[\hat{P}_{x_{i}},\hat{P}_{x_{j}}]=i\frac{\hbar^{2}}{l^{2}_{B}}\epsilon_{ij};~~[\hat{P}_{y_{i}},\hat{P}_{y_{j}}]=-i\frac{\hbar^{2}}{l^{2}_{B}}\epsilon_{ij},
\end{equation}
\begin{equation}\label{sym1} [\hat{x}_{i},\hat{P}_{x_{j}}]=[\hat{y}_{i},\hat{P}_{y_{j}}]=i\hbar\delta_{ij};
\end{equation}
It may be observed that in this case, neither the coordinates nor the momentum operators commute \cite{pnb2}. However, the operators 
\begin{equation}
    \hat{P}_{i}=\hat{P}_{x_i}+\hat{P}_{y_i}=\frac{eB}{c}\epsilon_{ij}(\hat{x}_{j}-\hat{y}_{j}),
    \label{jn}
\end{equation}
can act as  proper (commutative) translation generators, so that they satisfy the following commutation relations:
\begin{equation}
[\hat{x}_{i},\hat{x}_{j}]=il^{2}_{B}\epsilon_{ij};~~[\hat{P}_{i},\hat{P}_{j}]=0;~~[\hat{x}_{i},\hat{P}_{j}]=i\hbar\delta_{ij},
\label{nch}
\end{equation}
which represents a non-commutative Heisenberg algebra (NCHA) in two dimensions. In this instance, the operator-valued Hamiltonian of the effective system is given by
\begin{equation}
\hat{H}_{eff}=\frac{K_{0}}{2}(\hat{x}_{i}-\hat{y}_{i})^{2}+V(\hat{x}_{1})
\label{cv}
\end{equation}
A more conventional setting of this Hamiltonian in terms of the commutative translation generator $\hat{P}_i$ is as follows:
\begin{equation}
\hat{H}_{eff}=\frac{1}{2m_{B}}\hat{P}^{2}_{i}+V(\hat{x}_{1});~~~ i=1,2
\label{vv}
\end{equation}
where $m_{B}=\frac{e^2B^2}{c^2K_{0}}$ is the effective mass of the reduced two-particle system. It turns out to be instructive to introduce the pair of canonical variables:
\begin{equation}
\hat{R}_{i}=\frac {\hat{x}_{i}+\hat{y}_{i}}{2};~~\hat{P}_{i}=\frac{eB}{c}\epsilon_{ij}(\hat{x}_{j}-\hat{y}_{j});~~i, j=1, 2
\label{k1}
\end{equation}
where $\hat{R}_{i}$ is the centre of mass coordinate and $\hat{P}_{i}$ is the corresponding canonical momentum of our two-particle system. They satisfy the usual Heisenberg commutation relations as
\begin{equation}
[\hat{R}_{i},\hat{R}_{j}]=0;~~[\hat{P}_{i},\hat{P}_{j}]=0;~~~[\hat{R}_{i},\hat{P}_{j}]=i\hbar \delta_{ij}
\label{u}
\end{equation}
However, it is worth noting that the centre of mass position coordinates may also satisfy non-commutative Heisenberg algebra (NCHA) if the two particles are assumed to have different masses (For further details, see Appendix A). Even, if the two particles have the same mass, but their position coordinates satisfy NCHA with different non-commutativity parameters, in that case also, the centre of mass position coordinates can give rise to a non-commutative algebra.

Note further, that since the dynamics of the composite system is realized in terms of the coordinates of the positively charged particle, the information of the negatively charged particle is completely suppressed in the equations ($\ref{nch}$, $\ref{vv}$), but it is incorporated into the expression of commuting momentum operators. The extended Heisenberg algebra of the type as considered in eq.(\ref{nch}) has the important property: it is realizable in terms of commutative usual phase space variables (\ref{k1}) as
\begin{equation}
\hat{x}_{1}=Ad_{\hat U}(\hat{R}_{1});~~\hat{P}_{1}=Ad_{\hat U}(\hat{P}_{1})
\label{o}
\end{equation}
\begin{equation}
\hat{x}_{2}=Ad_{\hat U^\dagger}(\hat{R}_{2});~~\hat{P}_{2}=Ad_{\hat U^\dagger}(\hat{P}_{2})
\label{u}
\end{equation}
where we have made use of the fact of adjoint action: $Ad_{\hat{U}}(\hat{A})=\hat{U}\hat{A}\hat{U}^{\dagger}$ with a  quasi unitary operator $\hat{U}$:  
\begin{equation}
\hat U=\exp[{(-\frac{il^{2}_{B}}{2\hbar^{2}})\hat{P}_{1}\hat{P}_{2}}],
\label{uni}
\end{equation}
as it does not act unitarily on the entirely non-commutative phase space.

We can observe from the aforementioned equations ($\ref{o}$, $\ref{u}$) that the non-commutative phase space commutation algebra (\ref{nch}) can be simulated in terms of commutative phase space variables (canonical variables) i.e. the centre of mass coordinates as
\begin{equation}
\hat{x}_{i}=\hat{R}_{i}-\frac{c}{2eB}\epsilon_{ij}\hat{P}_{j}, ~~~~i,j=1,2
\end{equation}
It may be noted that this transformation is not canonical because it changes the commutation brackets. This transformation has occasionally been called a Darboux map \cite{n2} or Bopp's shift \cite{hm2}  which is of relevance  in the Bohmian interpretation of non-commutative quantum mechanics \cite{pbom}. Furthermore, this transformation with an explicit dependence on the deformation parameter, allows us to convert the Hamiltonian in NC space into a modified Hamiltonian in commutative equivalent space. It follows that if we are able to solve the spectrum of the system Hamiltonian in commutative equivalent space, we can also  obtain the spectrum of the system in primitive non-commutative space, though the states in both situations are not the same. We will discuss how the aforementioned maps aid in the extraction of non-classical cat states in the next section.

\section{Preparation of Schr\"{o}dinger Cat states}

Using the formalism presented in the previous section, we are now in a position to investigate the main goal of this work, {\it viz.}, how we might naturally prepare Schr\"{o}dinger's Cat states. To do so, we first  consider a particular Hamiltonian with a harmonic oscillator potential in the $\hat{x}_1$ direction, given by
\begin{equation}
\hat H_{eff}\rightarrow\hat{H}_{NC}=\frac{\hat{P}^{2}_{1}}{2m_{B}}+\frac{\hat{P}^{2}_{2}}{2m_{B}}+V(\hat{x}_{1}),
\label{mh}
\end{equation}
where $V(\hat{x}_{1})=\frac{1}{2}K\hat{x}^{2}_{1}$ and $m_{B}=\frac{e^2B^2}{c^2K_{0}}$.
The corresponding time dependent Schr\"{o}dinger equation is:
\begin{equation}
    i\hbar\frac{\partial}{\partial t}\ket{\psi(t)}_{NC}=\hat{H}_{NC}\ket{\psi(t)}_{NC}
    \label{sch}
\end{equation}


Furthermore, because of the non-commutativity of this theory, it is impossible to construct simultaneous eigenstates with noncommutative coordinates, which makes it difficult to define a local probability density for the wave-function that corresponds to a particular state $|\psi(t)>_{NC}$ \cite{pbom1}. However, this issue can be bypassed by using the interpretation mentioned in \cite{pbom1}, or by using the coherent states formulation of noncommutative quantum mechanics with the help of the Voros product \cite{g}. 

In our present case, it can be easily observed that the system  Hamiltonian mentioned above can be rewritten  as,
\begin{equation}
\hat {H}_{NC}=\hat{U}\hat{H}_{CM}\hat{U}^{\dagger},
\end{equation}
with
\begin{equation}
  \hat{H}_{CM}=\frac{\hat{P}^{2}_{1}}{2m_{B}}+\frac{\hat{P}^{2}_{2}}{2m_{B}}+V(\hat{R}_{1}),  
\end{equation}
where we have used the fact that $V(\hat{x}_{1})=V(\hat{U}\hat{R}_{1}\hat{U}^{\dagger})=\hat{U}V(\hat{R}_{1})\hat{U}^{\dagger}$. Here $\hat{H}_{CM}$ is the unitarily equivalent form of the system Hamiltonian expressed in terms of the Center of Mass coordinates, whereas the $\hat{H}_{NC}$  represents the system Hamiltonian written in terms of the positively charged particle coordinates. We can readily recognize that $V(\hat {R}_{1})=\frac{1}{2}K\hat{R}^{2}_{1}$, where $K$ is the spring constant of the impurity interaction faced by the positive charge in the $\hat{x}_{1}$ direction only. 
Accordingly, the Schr\"{o}dinger equation ($\ref{sch}$) transforms as follows:
\begin{equation}
i\hbar\frac{\partial}{\partial t}\ket{\psi(t)}_{CM}=\hat{H}_{CM}\ket{\psi(t)}_{CM}    
\end{equation}
where $\ket{\psi(t)}_{CM}= \hat{U}^{\dagger}\ket{\psi(t)}_{NC}.$

To describe the natural generation of entangled cat states, we consider the appropriate eigenstates of the unitarily equivalent Hamiltonian ($\hat{H}_{CM}$), now presented as:
\begin{equation}
\ket{\psi_{0}}_{CM}=\ket{0}\otimes[d_{+}\ket{+k_{2}}+d_{-}\ket{-k_{2}}],
\end{equation}
where $|d_{+}|^2$ and $|d_{-}|^2$ denote the probability of finding the free particle with nonzero momentum in $\ket{+k_{2}}$ and $\ket{-k_{2}}$ states, respectively, and $\ket{0}$ represents the ground state of the 1D harmonic oscillator system with $\hat{a}_{1}$ and ${\hat{a}_{1}}^\dagger$ representing the corresponding annihilation and creation operators, respectively, satisfying the following algebra:
\begin{equation}
[\hat{a}_{1},\hat{a}^{\dagger}_{1}]=\mathbb{I};~~~~\hat{a}_{1}=\frac{m_{B}\omega_{B}\hat{R}_{1}+i\hat{P}_{1}}{\sqrt{2m_{B}\omega_{B}\hbar}};~~~~~\hat{a}_{1}\ket{0}=0,
\end{equation}
with $\omega_{B}=\sqrt{\frac{K}{m_{B}}}$, and $\ket{\pm k_{2}}$ corresponds to the right and left moving free particle's momentum states, respectively, which satisfy:
\begin{equation}
\hat{P}_{2}\ket{\pm k_{2}}=\pm P_{2}\ket{\pm k_{2}};~~~~P_{2}=\hbar k_{2}
\end{equation}
The state vector corresponding to the non-commutative phase space (or in terms of the positively charged particle coordinates) is given by
\begin{equation}
\ket{\psi_{0}}_{NC}=\hat U\ket{\psi_{0}}_{CM},
\label{sc}
\end{equation}
 $\ket{\psi_{0}}_{NC}$ can be expressed as, 
\begin{align}
\ket{\psi_{0}}_{NC}=(\exp[(-\frac{il^{2}_{B}}{2\hbar^{2}})\hat{P}_{1}\otimes\hat{P}_{2}])\notag\\
  [\ket{0}\otimes(d_{+}\ket{+k_{2}}+d_{-}\ket{-k_{2}})]
\end{align}
which leads to
\begin{align}
\ket{\psi_{0}}_{NC}=d_{+}([\exp(-\frac{il^{2}_{B} k_{2}}{2\hbar}\hat{P}_{1})]\ket{0})\otimes\ket{+k_{2}}\notag\\+d_{-}([\exp(\frac{il^{2}_{B} k_{2}}{2\hbar}\hat{P}_{1})]\ket{0})\otimes\ket{-k_{2}}
\end{align}
On substituting $l^{2}_{B}=\frac{\hbar c}{eB}$ in the above equation, we arrive at:
\begin{align}
\ket{\psi_{0}}_{NC}=d_{+}([\exp[{(-i\frac{ck_{2}}{2eB})}\hat{P}_{1}]]\ket{0})\otimes\ket{+k_{2}}\notag\\
+d_{-}([\exp[{(i\frac{ck_{2}}{2eB})\hat{P}_{1}}]]\ket{0})\otimes\ket{-k_{2}}
\end{align}

Now, for a harmonic oscillator potential, the momentum operator $\hat{P}_{1}$ can be written  as-
\begin{equation}
\hat{P}_{1}=i\sqrt\frac{m_{B}\omega_{B}\hbar}{2}({\hat{a}_{1}}^\dagger-\hat{a}_{1})
\end{equation}
Putting the above expression in equation (33), we obtain,
\begin{align}
\scalebox{0.8}{$\ket{\psi_{0}}_{NC}=d_{+}([\exp[{(\frac{ck_{2}}{2eB})}\sqrt\frac{m_{B}\omega_{B}\hbar}{2}({\hat{a}_{1}}^\dagger-\hat{a}_{1})]]\ket{0})\otimes\ket{+k_{2}}$}
\notag\\
          \scalebox{0.8}{$ +d_{-}([\exp[{(-\frac{ck_{2}}{2eB})}\sqrt\frac{m_{B}\omega_{B}\hbar}{2}({\hat{a}_{1}}^\dagger-\hat{a}_{1})]]\ket{0})\otimes\ket{-k_{2}}$}
           \label{ncs}
\end{align}
It follows that the above state vector ($\ref{ncs}$)  may also be written in the form of a superposition of single-component coherent states as
\begin{align}
\scalebox{0.8}{$\ket{\psi_0}_{NC}=d_{+}\ket{+\alpha}\otimes\ket{+k_{2}}+d_{-}\ket{-\alpha}\otimes\ket{-k_{2}}$},
\label{m}
\end{align}
wherein  $\ket{\pm \alpha}=e^{\pm\alpha({\hat{a}_{1}}^\dagger-\hat{a}_{1})}\ket{0}$ with $\alpha=\frac{ck_{2}}{2eB}\sqrt\frac{m_{B}\omega_{B}\hbar}{2}$ are real-valued coherent states (or a displacement of the vacuum) that belong to the subset of the over complete space of usual complex parameter valued coherent states \cite{bom1}.\\


Here it may be worthwhile to mention a  property of the coherent state $\ket{\pm \alpha}$: the dimensionless parameter $\alpha$ may be rewritten as
\begin{equation}
  \alpha=\frac{1}{2}P_{2}({\frac{K}{K_{0}}})^{1/4}  \sqrt{\frac{c}{2eB\hbar}}=\xi k_{2} l_{B}
 \label{lk}
\end{equation}
with $\xi=\frac{1}{2}(\frac{K}{4K_{0}})^{\frac{1}{4}}$ .
A coherent state $\ket{\alpha}$ can have an arbitrarily large amplitude, and
hence, the energy of a macroscopic harmonic oscillator \cite{scv} can be approximated by the energy of a one-dimensional quantum mechanical HO by
suitably choosing $\mid \alpha\mid$ to be arbitrarily large. For large enough $\mid\alpha\mid$ values,  $\ket{+\alpha}$ and $\ket{-\alpha}$ correspond to macroscopically distinguishable states and may be labelled as `(+) (alive)' and `(-) (dead)' \cite{sv,gl}. In this sense, we can regard the above state (\ref{m}) as an entangled SCS, 
holding $\mid\alpha\mid \sqrt{h}$ fixed with finite value in the classical limit \cite {v,s}. Accordingly, one may
consider $\ket{\pm\alpha}$ to be ``classical-like'' states, but their coherent superposition is endowed with non-classical properties. A similar type of Schr\"{o}dinger cat state has been explored in the context of Rydberg atomic systems using pulsed signals \cite{vn}, where the coherent (classical) states are coupled with the internal spin states of the atom. However, in our present case, the coherent states are coupled with the left and right-moving free particle states.

In the primitive non-commutative phase space, we may  rewrite the state vectors (\ref{ncs})
in the following concise way:
\begin{align}
\ket{\psi_{0}}_{NC}\rightarrow  \kittyket= \mathcal{N}[\ket{+\alpha;+k_{2}}+e^{i\phi}\ket{-\alpha;-k_{2}}];~~\notag\\
\ket{\pm\alpha;\pm k_{2}}=\ket{\pm\alpha}\otimes \ket{\pm k_{2}},
\label{scat}
\end{align}
with an arbitrary phase factor ($\phi$) and normalization constant $\mathcal{N}$. For the aforementioned reason, the states $\ket{\pm \alpha}$ may be considered to be ``macroscopic" like states with the same amplitude but opposite in phase. (in the present case, the $\mid\alpha\mid$ parameter is not arbitrary, but is defined in terms of the spring constants, magnetic field
and electric charge). However, their superposition (\ref{scat}) has several non-classical characteristics \cite{op}. Particularly, for the relative phase factor $e^{i\phi}=\pm 1$, we get even and odd cat states that have been well-studied in the literature \cite{2,bose}. Moreover, it is evident from (\ref{scat}) that the coherent states and the free particle states are  entangled: when the coherent state parameter has a positive sign, the free particle state is right-moving. On the other hand, the free particle state
is left-moving when the coherent state parameter has a negative sign. Therefore, $\ket{\psi_{0}}_{NC}$ is an entangled Schr\"{o}dinger cat state 
containing  the coherent superposition \cite{cohsup1, cohsup2} of two states that are diametrically opposite to one another. 

It may be emphasized that in our  model, the free particle-like nature along the $\hat{x}_{2}$ direction plays a pivotal role in generating left-right superposition states. These states can entangle with coherent states to form cat-like states. A specific superposition of free particle states can also form a localized free particle state (wave packet), exhibiting multi-component cat states. In this regard, it is important to emphasize that the free particle nature in the $\hat{x}_{2}$ direction of our model is only an effective description of the system under very strong magnetic field conditions. In solid-state physics, the concept of free particles often serves as a starting point to describe more complex systems. Specifically, in semiconductors, the effective mass approximation allows us to treat electrons in the conduction band and holes in the valence band as free particles. Near the bottom of the conduction band or the top of the valence band, the energy-momentum relationship can be approximated by a parabolic dispersion relation, similar to that of a free particle. These are often referred to as “quasi-free" particles, simplifying the analysis of electronic and optical properties.

Since a momentum eigenstate is an idealization \cite{kop},  we consider a more 
realistic scenario in which the system's motion in the commutative phase space is  localized within a specific length scale $\sigma$ along the $\hat{R}_2$ direction. In this case, we generalize the notion of free particle states to a propagating Gaussian state given by
\begin{equation}
   \ket{\psi_{G}}=\sqrt{\frac{\sigma}{\sqrt{\pi}}}\int_{-\infty}^{+\infty} e^{-\frac{\sigma^{2}}{2}(k_{2}-k_{0})^{2}} \ket{k_{2}} dk_{2}
\end{equation}
where $\sigma$ is the width and $k_{0}$ is  the peak momentum of the wave packet.
Now, following the prescription of  (28), we can write the composite state of the particle, when the dynamics of the system are realized in terms of the centre of mass coordinates, as
\begin{equation}
    \ket{\psi_{0}}_{CM}=\ket{0}\otimes\ket{\psi_{G}}
    \label{go}
\end{equation}
Accordingly, we can generalize the notion of a two-component cat state (\ref{scat}) to
\begin{align}
\scalebox{0.8}{$  \kittyket=\hat{U}\ket{\psi}_{CM}=\sqrt{\frac{\sigma}{\sqrt{\pi}}}\int^{+\infty}_{-\infty}\ket{\alpha(k_{2})} \otimes \ket{k_{2}}e^{-\frac{\sigma^{2}}{2}(k_{2}-k_{0})^{2}} dk_{2} $}
    \label{mca}
\end{align}
which describes a multi-component entangled Schr\"{o}dinger cat state \cite{rop1} where each component is specified through the momentum eigenvalues. Such
a state is highly non-classical, which can be verified through the 
corresponding Wigner function  \cite{rop1}. Thus, in the presence of a strong magnetic field background, one may successfully prepare a Schr\"{o}dinger Cat State utilizing a non-relativistic electric dipole model, where non-commutativity plays an important role. It may be reiterated here that we 
explore the system in terms of the positively charged particle coordinates.

Note that, we specifically consider the impurity potential attached to the positively charged particle to depend solely on one direction, $\hat{x}_{1}$. This is crucial for achieving our desired outcome. Introducing a similar type of potential (harmonic oscillator type) along the $\hat{x}_{2}$ direction would bring the non-commutative nature of the two components, $[\hat{x}_{1}, \hat{x}_{2}] \neq 0$, into effect. Transforming the ground state $|\Psi_g\rangle_{CM}$ to $\hat{U} |\Psi_g\rangle_{CM}$ with the unitary operator $\hat{U}$ would result in a superposition of displaced number states rather than coherent states. This transformation disrupts the coherent superposition of states necessary for realizing cat-like states. Therefore, we consider the potential as a function of only $\hat{x}_{1}$ to naturally demonstrate the emergence of cat-like states.
Similarly, if we attach the potential along $\hat{y}_{1}$ with the negatively charged particle, the potential term affects the dynamics of the first charged particle in the presence of the potential attached to the positively charged particle along the $\hat{x}_{1}$ direction. This effectively introduces a coupling between the two modes of the positively charged particles with respect to the center of mass frame, resulting in the loss of the coherent superposition of coherent states, thus preventing the formation of cat-like states. This indicates that achieving cat-like states is possible only for a specific class of potentials, which should be attached to either the positively charged particle or the negatively charged particle in a single direction, rather than applying potentials to both charged particles in both directions. For more details, see Appendix C.

\section{Collapse and revival of entanglement of SCS}
In this section, we will begin by investigating the degree of entanglement of the SCS state $\ket{\psi}_{NC}$. In order to do so, we first write down the
corresponding density matrix given by
\begin{align}
    \hat{\rho}_{NC}=(\sqrt{\frac{\sigma}{\sqrt{\pi}}})^2\int_{-\infty}^{+\infty}
    \int_{-\infty}^{+\infty}[ \ket{\alpha(k_{2})}_A\bra{\alpha(k_{2}^{\prime})}] \notag\\
    \otimes [\ket{k_2}_B\bra{k_2^{\prime}}]e^{\frac{-\sigma^2}{2}(k_2-k_0)^2} e^{\frac{-\sigma^2}{2}(k_2^{\prime}-k_0)^2} dk_2 dk_2^{\prime}
    \label{k}
\end{align}
where the subscripts $A$ and $B$ denote two distinct subsections of our bipartite system, one of which is associated with coherent states and the other with momentum eigenstates, each of which corresponds to two distinct degrees of freedom in the non-commutative plane. Since $|\psi>_{NC}$ is a composite pure state, the entanglement
between the coherent states and free particle states can be
quantified in terms of the von-Neumann entropy  given by
\begin{equation}
S=-Tr_{A}[\hat{\rho}_{red}~ln (\hat{\rho}_{red})] 
\end{equation}
where the reduced density matrix is defined as 
\begin{align}
\label{reduced}
    \hat{\rho}_{red}=\text{Tr}_B [\hat{\rho}_{NC}]\notag\\
    =\frac{\sigma}{\sqrt{\pi}}\int_{-\infty}^{+\infty}[\ket{\alpha(k_2)}_A\bra{\alpha(k_2)}]e^{-\sigma^2(k_2-k_0)^2} dk_2
\end{align}
with
\begin{equation}
    \text{Tr}(\hat{\rho}_{red})=\frac{\sigma}{\sqrt{\pi}}\int_{-\infty}^{+\infty}e^{-\sigma^2(k_2-k_0)^2}dk_2=1
\end{equation}
For the present purpose, it suffices to compute the purity function \cite{purity}, given by
\begin{align}
\label{purityfn}
    \text{P}(\alpha)=\text{Tr}(\hat{\rho}^2_{red})=\sum_n \bra{n}\hat{\rho}^2_{red}\ket{n}\notag\\
    =\sum_m\sum_n \bra{n}\hat{\rho}_{red}\ket{m}\bra{m}\hat{\rho}_{red}\ket{n}
\end{align}
After a little algebra, one obtains
\begin{align}
\label{59}
    \bra{n}\hat{\rho}_{red}\ket{m}=\frac{\sigma}{\sqrt{\xi^{2}l_{B}^2+\sigma^2}}\frac{1}{\sqrt{n!}\sqrt{m!}} e^{(-\sigma^2k_0^2)}\notag\\
    (\frac{\xi l_{B}}{2\sigma^2})^{n+m}\frac{\partial^{n+m}}{\partial k^{n+m}_0}(e^{\frac{\sigma^4k_0^2}{\xi^{2} l_{B}^2+\sigma^2}})
\end{align}
By inserting equation (\ref{59}) into (\ref{purityfn}) it follows that
\begin{align}
\label{45}
    \scalebox{0.9}{$\text{P}(\xi_{0};l_{B})=(\frac{1}{1+\xi^{2}_{0}})e^{(-2\sigma^2 k_0^2)} [e^{(\frac{\sigma^2k_0^2}{1+\xi_{0}^{2}})} (e^{\frac{\xi^{2}_{0}}{2\sigma^2}\overset{\leftarrow}{\frac{\partial}{\partial k_0}} \overset{\rightarrow}{\frac{\partial}{\partial k_0}}})  e^{(\frac{\sigma^2k_0^2}{1+\xi_{0}^{2}})}]$};
\end{align}
where $\xi_{0}=\frac{\xi l_{B}}{\sigma}$ and $\xi=\frac{1}{2}\left(\frac{K}{4K_{0}}\right)^{\frac{1}{4}}$ as defined in equation (\ref{lk}).
The above expression  can be rewritten (see Appendix B) as
\begin{equation}
 \text{P}(\xi_{0}; l_{B})=\frac{1}{\sqrt{{1+2\xi^{2}_{0}}}}  
\end{equation}
In Figure 1, we plot the purity function versus the parameter $\xi_{0}$. It can be observed
that the purity function reduces from unit value (separable or disentangled state) with
increase of the parameter $\xi_{0}$, indicating increment of entanglement in the
system for higher values of $\xi_{0}$ (or lower values of the width of
the wave packet $\sigma$). We consider the quantum length scale $l_B=1.483 \times10^{-8}$m, and vary the width of the wave-packet in the range of $O(10^{-11}\to 10^{-6})$. Different $l_B$ values displayed in the figure may
originate due to the variation of the magnetic length scale with different accessible magnetic fields in the laboratory.

\begin{figure}[h]
    \centering
    \includegraphics[scale=0.85]
    {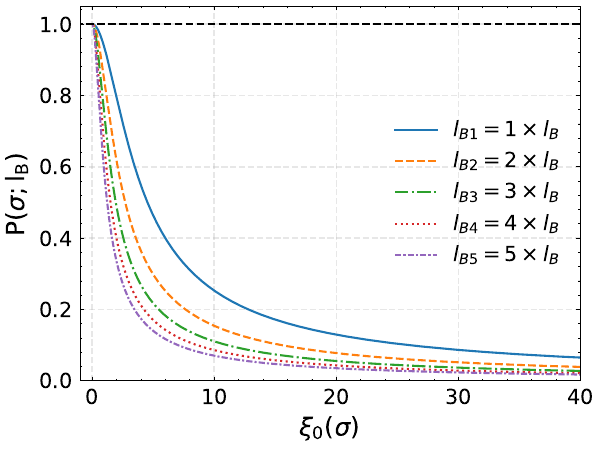}
    \caption{The Purity function is plotted against the dimensionless factor $\xi_0$ which varies inversely with the width of the wave-packet $\sigma$. Plots for several choices of the quantum length scale are displayed and the direction of the arrow suggests the direction of increment of magnetic fields, which is inversely proportional to the $l^{2}_B$.}
    \label{fig:enter-label}
\end{figure}

It may be noted that if we just assume $\xi_{0}<<1$ with $\xi \sim 1$  which implies that $l_{B}<<\sigma$, i.e, the width of the Gaussian packet ($\sigma$) is large enough compared to the magnetic quantum length scale such that  we can ignore $\xi_{0}$, then  it leads to the unit value of the purity function, or in other words, the collapse of the entanglement 
in the state. On the other hand, we can make the states entangled by choosing $\sigma$ comparable to the magnetic length scale $l_B$ where $\text{P}(\xi_{0};B)$ becomes less than unity. More interestingly, the revival of the entanglement state can occur,  if one considers a time-dependent regime. Let us recall from the definition of $\xi$, that it basically depends on the coupling  strength $K$ of the ``impurity" interaction.

The dynamic behaviour of impurities in materials is known to lead to
time-varying spring interaction \cite{Dj1,Dj6}. Such a dynamic nature of the
coupling has been studied in the literature in the context of several physical
systems such as in optical lattices \cite{Dj5}, and extensively in the domain 
of quantum electronic transport \cite{Dj,Dj3,Dj4}. Let us now consider that the spring ``constant" $K$ is a slowly varying periodic function of time due to some external effects, with the time-variation given by 
\begin{equation}
K  \rightarrow K(t)=K \text{cos}^{4}\omega_{d}t=K \text{cos}^{4}\theta(t)
\end{equation}
which clearly indicates 
$\xi(t)=\frac{1}{2}(\frac{K \text{cos}^{4}\omega_{d} t}{4K_0})^{(1/4)}$  and $\xi_{0}(t)=\frac{\xi(t) l_{B}}{\sigma}$. Hence, the purity function  gets modified to:
 \begin{equation}
 \label{purityperiodic}
 \text{P}(\xi_{0}; l_{B})=\frac{1}{\sqrt{1+2{\xi^{2}_{0}(t)}}}
\end{equation}
Before we proceed further, the following comments are in order.
It may be noted that the expression above ensures that the purity function (\ref{purityperiodic}) remains real  at any arbitrary instant $t$ within the time period. Considering a simple form $K(t) = K \cos^n(\omega_{d}t)$ with $\xi(t=0)\sim1$ to highlight the main features, we observe that $\xi_{0}(t)\sim \cos^{\frac{n}{4}}(\omega_{d}t)$. The reality condition of the purity function is satisfied only when we limit ourselves to $n/2$ being an even integer. For the purpose of highlighting the main features of the dynamical behavior of entanglement, we choose the smallest integer $n=4$. The periodic behaviour of the
purity function gives rise to the phenomenon of collapse and revival of
entanglement, as we now show.

It should be noted that even when $\sigma$ is comparable to the magnetic length scale $l_B$, disentanglement occurs in specific instances such as $t_{d}=\frac{\pi}{2\omega_d},\frac{3\pi}{2\omega_d},\frac{5\pi}{2\omega_d},\frac{7\pi}{2\omega_d},\ldots$ with a separation of the time period $\frac{\pi}{\omega_d}$ between two successive collapses. For the rest of the time interval, the states are entangled. This distinguishing feature is known as the collapse and revival of entanglement in the literature \cite{rop2}. In Figure 2, we plot the purity function versus
the periodic parameter $\theta(t)$ for several different values of the width of the wave packet $\sigma$. It is clearly seen that the magnitude of entanglement
revival increases more for narrower wave-packets.

\begin{figure}[h]
    \centering
    \includegraphics[scale=0.82]{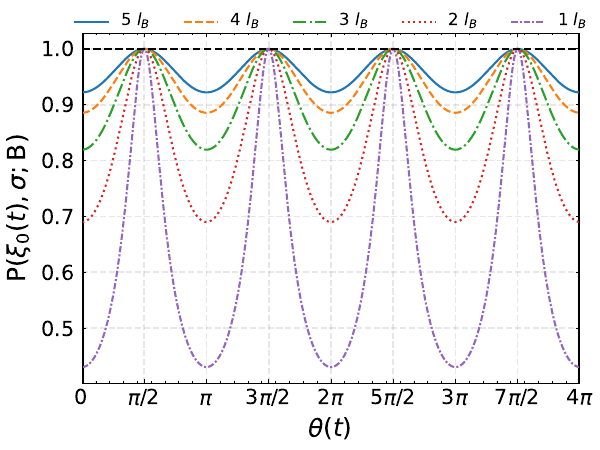}
    \caption{Time evolution of the Purity function is plotted against the parameter $\theta(t)$, for various 
widths of the wavepackets $\sigma$.}
    \label{fig:enter-label}
\end{figure}

Here, it needs to be mentioned that in order to observe entanglement revival of the states, it is required to choose $\sigma$ of the order comparable to that of the magnetic length scale $l_B$ or less, as $-1\leq \text{cos}\omega_d t\leq+1$. On the other hand, if we choose  $\sigma$ to be much larger than $l_B$, then the additional term in the denominator of Eq.(\ref{purityperiodic}) can be completely negligible which will take us back again to the situation of the entanglement collapse, {\it viz.} $\text{P}(\xi_{0}; B) \sim 1$. Instances of the phenomenon of entanglement collapse and revival have been pointed out earlier in the literature predominantly in
the context of the Jaynes-Cummings model for optical systems \cite{R2, rop2}. Here we furnish a striking example of entanglement collapse and revival in the context of an excitonic dipole in a condensed matter system.

To illustrate the collapse and revival of entanglement in cat states with moving free particle states, we have used a time-dependent spring constant \( K(t) \) within our dipole system. The effective masses of electrons and holes, denoted by \( m \), are taken to be of the same order of magnitude,
which allows the center-of-mass (CM) coordinates to be treated as commutative with respect to the positive charge particle coordinates \cite{exp}. Though the effective masses of electrons and holes generally differ due to the unequal curvature of the valence and conduction bands, it is possible
for them to be approximately equal if there exists electron-hole symmetry
manifested by the valence band maxima aligning with the conduction band minima at the same point. The experimental realization of our model
may be feasible in systems involving certain specific direct band-gap semiconductors \cite{exciband2, exciband3, exciband1}, where the minimum conduction band for electrons and the maximum valence band for holes are located at the same point of the Brillouin zone. In these materials, the effective masses of electrons and holes can be quite similar in magnitude, typically arising from specific band structures and symmetries.

\begin{figure}[h]
    \centering
    \includegraphics[scale=0.9]{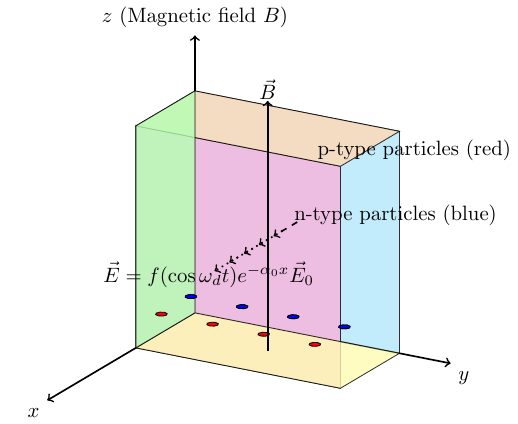}
    \caption{This figure illustrates a 3D view of a semiconductor layer with p-type and n-type particles. The axes are labeled \(x\), \(y\), and \(z\). The semiconductor layer is shown with different colors: green for the left face, cyan for the right face, yellow for the bottom face, orange for the top face, and magenta for the front face. The p-type particles are shown as red circles on the front layer, and the n-type particles are shown as blue circles inside the bulk. The electric field is represented by dashed and dotted arrows pointing in the \(x\)-direction, while the magnetic field is shown by a solid arrow along the \(z\)-axis. The electric field is given by \(\vec{E} = f(\cos \omega_{d} t) e^{-\alpha_{0} x} \vec{E}_{0}\), and the magnetic field is denoted by \(\vec{B}\).}
    \label{fig:enter-label}
\end{figure}

A possible realization of our model is illustrated in Figure 4, involving a time-varying electric field \(\vec{E} = f(\cos \omega_{d} t) e^{-\alpha_{0} x} \vec{E}_{0}\) applied along the x-axis, perpendicular to the surface layer where the holes are located \cite{exc}. In fact, the holes and electrons, which are confined in the bulk, reside in a plane parallel to the x-y plane, while a magnetic field is applied along the z-axis. This setup is ideal for a heavily doped p-type semiconductor, where an accumulation layer of holes is formed near the surface \cite{exciband2,exciband5}.
The AC electric field varies slowly (to avoid abrupt changes to the initial state, specifically the ground state of the oscillator trap, which is crucial for demonstrating cat-like states), with \( e^{-\alpha_{0} x} \) ensuring that its primary effect is near the surface. Here, due to the high conductivity of heavily doped p-type semiconductors, 
 $\alpha_{0}^{-1}$  can be approximated by
$\alpha_{0}^{-1} \sim \sqrt{\frac{2}{\sigma \mu \omega_{d}}},$ (using the approximations of a good conducting medium), where \(\sigma\), \(n_{h}\), and \(\mu\) denote conductivity, hole doping concentration, and permeability of the conducting medium, respectively. For this good conductivity of the p-type sample, even low-frequency fields will have a limited penetration depth $\alpha_{0}^{-1}$, thus restricting their effect on the bulk material. This limited penetration depth ensures that the field primarily influences surface-localized holes, which aligns with our toy version of the excitonic dipole model. The electric field introduces a time-dependent potential that mimics a time-dependent spring constant \( K(t) \) in the Hamiltonian, under suitable attenuation and approximations. By employing \( \cos^4(\omega_d t) \), we ensure that the purity function remains real, as discussed after equation (\ref{purityperiodic}). In fact, any periodic time-dependent function that maintains an oscillatory nature at \( t=0 \) (i.e., has a non-zero value at \( t=0 \)) and satisfies the reality condition of the purity function can serve as the time-dependent spring constant \( K(t) \). This allows us to explore phenomena such as the variation of the purity function, the revival and collapse of entanglement, and other related effects.

Before concluding, it may be noted that we have considered a periodic function of time with a slow variation, characterized by a time period $t_{d}=\frac{2\pi}{\omega_{d}}$. The degree of adiabaticity, which quantifies the slowness of the process, can be defined as
\begin{equation}
\epsilon=\hbar\frac{\mid\bra{m(t)}\frac{\partial \hat{H}(t)}{\partial t}\ket{n(t)}\mid}{(E_{m}-E_{n})^{2}}|\sim \frac{\omega_{d}}{\omega_{mn}},
\end{equation}
where the transitional angular frequency $\omega_{mn}=\frac{E_{m}-E_{n}}{\hbar}$ is introduced. $\epsilon << 1$ signifies that the variation of the Hamiltonian (in terms of matrix elements) over the time scale $\frac{2\pi}{\omega_{nm}}$ is small compared to the energy separation $E_{m}-E_{n}$. The adiabatic approximation is valid when this condition is satisfied, serving as a measure of the slowness required for its applicability \cite{o2}. The dynamical periodic behavior of the purity function is a natural consequence of a harmonic oscillator system with a periodically time-varying spring constant in an adiabatic process. In this context, the system can adjust or adopt itself to stay in its initial eigenstates, preserving the real forms of the purity function. Though the adiabatic approach has been invoked here to simplify the technical complexities and extract the essential features of the dynamic nature of entanglement,  non-adiabatic time variation may also lead to collaps revival  behavior\cite{o3}.

\section{Conclusions}

To summarize, in this work, we have considered a composite two-particle planar dipole system in the presence of a strong constant and uniform magnetic field, in which two oppositely charged particles interact via harmonic interaction, in addition to an impurity interaction  experienced by the positively charged particle.  Our system may be regarded as a toy version of  excitonic dipole models that can be realized in  some specific direct band gap semiconductors  \cite{exciband2, exciband3, exciband1} having the conduction band minimum  for electrons and the valence band maximum  for holes   both located at the same point of the Brillouin zone, where the effective mass of electrons and holes can be quite similar in magnitude. This typically arises due to specific band structures and symmetries of materials.
The additional interaction could arise from intrinsic features such as defects or impurities, as well as from external influences like an external electric field or strain in the material \cite{e4}.

In our analysis, we have first addressed the classical picture  in the
context of our system's Lagrangian formulation which is the most natural in a strong magnetic field limit. Using symplectic analysis of this first-order Lagrangian, we have specified the canonical/Weyl-Moyal type deformed NC classical phase space to be an intrinsic part of our model. Next, we have explored the quantum mechanical description of our model by elevating all the phase space variables to the level of Hermitian operators. The spatial and momentum sectors of individual charged particles obey a non-commutative deformed algebra. Here, the non-commutativity emerges as a natural consequence of placing two oppositely charged particles in a strong constant background magnetic field. The square of the magnetic length scale acts as the effective non-commutative parameter. 

We have presented a physical interpretation of the mapping from the deformed phase space to the usual commutative phase space. The non-commutative phase space represents the system Hamiltonian written in terms of the positively charged particle coordinates, while the standard quantum mechanical phase space is more suitable for describing our system in terms of the composite system's centre of mass coordinates. The dynamics can, therefore, be analyzed in terms of non-commuting variables or, alternatively, using phase space transformations, in terms of commuting variables. In literature,  non-commutativity has been often introduced by hand for a single point particle, thus ruling out any  physicality of commutative phase-space variables in such cases. However, in the present case, non-commutativity emerges naturally, thereby giving a physical meaning to the commutative phase-space variables. Determining the Hamiltonian's ground state in the commutative phase space allows us to express the quantum state in the non-commutative phase space  as a  superposition of two diametrically opposite coherent states, entangled with momentum eigenstates. This reveals the emergence of entangled and two-component as well as multi-component Schr\"{o}dinger Cat States (SCS) in our system.
 
Furthermore, we have estimated the magnitude of entanglement in the system of multicomponent entangled cat states.  By utilizing the purity function, we  demonstrate that the effective non-commutative parameter ($l_B^2$) is responsible for the entanglement. We show that when the width of the Gaussian wave packet ($\sigma$) significantly exceeds the minimal length scale ($l_B$), the entangled cat states undergo collapse. Conversely, when $\sigma$ is comparable to the nonzero magnetic length scale $l_B$, the entanglement can be observed. Moreover, we show that if time-dependent impurity potential is
chosen, entanglement revival and collapse occurs periodically. So notably, within the same formalism, we observe the phenomenon of collapse and revival of entanglement in the non-commutative plane in the time-dependent regime with a suitable choice of the $\sigma$ parameter for the revival case, while the collapse is completely controlled by the nodes of the periodic function involved in the impurity interaction.

Our study explores the natural emergence of Schr\"{o}dinger cat states in excitonic models, where these states can be conceptualized as qubits \cite{ep}. Specifically, the states are represented as superpositions:
$
|+\alpha(k_{2})\rangle \otimes |+k_{2}\rangle \equiv |1\rangle_{k_{2}}$ and
$
|-\alpha(k_{2})\rangle \otimes |-k_{2}\rangle \equiv |0\rangle_{k_{2}}$, where the momentum eigenstates are distinct and isolated, {\it i.e.}, having zero overlap between $|+k_{2}\rangle$ and $|-k_{2}\rangle$.
Our approach  indicates that multiple components of cat states, which represent superpositions of large qubit collections with fixed momentum eigenvalues, could be used to encode qubits for constructing extensive quantum registers \cite{ps}. Many-body effects in excitonic Bose-Einstein condensates can lead to Schr\"{o}dinger cat states, potentially resulting in new phases of matter at critical temperatures. Additionally, our system has potential applications in quantum error correction by encoding each logical qubit using multimode cat states \cite{qerror1, qerror2}. Experimentation of our model proposing the generation of excitonic cat states can pave a new path for studying the emerging field of optoelectronics exploring fundamental physics of quantum entanglement.

Finally, it may be noted that spin-orbit interactions in solid-state systems introduce electronic band curvature, leading to the emergence of Berry curvature in momentum space. Such Berry curvature modifies the usual phase space symplectic structure of Bloch electrons \cite{xio,xio2}. In light of non-commutative quantum mechanics,  our present analysis can be extended to include investigations on the possible emergence of Schr\"{o}dinger cat states in solid state systems involving the 2D excitonic Coulomb problem with the Berry curvature of the electron's and the hole's Bloch states \cite{Be1,Be2,Be3}.  This extension opens a gateway to a new frontier, promising fresh insights and potentially paving the way for the experimental exploration of the ``macroscopic" quantum state of excitons \cite{sp,ps}.

\section{Acknowledgements}

PN and ND acknowledge the support of S.N. Bose National Centre for Basic Sciences, where this work was initiated. PN also thanks the
Department of Physics, Stellenbosch University, for providing postdoctoral funds during the period when a major part of this work was completed.  The authors thank Biswajit Chakraborty, Debasish Chatterjee, Ananda Dasgupta, and Frederik G. Scholtz for some fruitful discussions. ASM acknowledges support from the Project No. DST/ICPS/QuEST/2018/98 from the Department
of Science and Technology, Government of India. The authors thank the referees for valuable comments and
suggestions.

\section{Appendix A}
Here we present a manifestation of the non-commutativity of the centre of mass coordinates arising in the case of two oppositely charged particles with different masses $m_{+}$ and $m_{-}$ representing the masses of positive and negatively charged particles respectively.
The corresponding centre of mass (CM) coordinates of the above-discussed system is-
\begin{align}
\hat{R}_i=\frac{m_{+}\hat{x}_i+m_{-}\hat{y}_i}{m_{+}+m_{-}};\notag\\
~\hat{P}_{i}=\hat{P}_{x_i}+\hat{P}_{y_i}=\frac{eB}{c}\epsilon_{ij}(\hat{x}_{j}-\hat{y}_{j});~~i, j=1,2
\end{align}
Now, utilizing the results obtained from equation (10), the commutation brackets between the CM coordinates can be obtained in the following form-
\begin{equation}
[\hat{R}_i,\hat{R}_j]=\frac{m_{+}^2-m_{-}^2}{(m_{+}+m_{-})^2}il_B^2\epsilon_{ij};~~i,j=1,2 
\end{equation}
clearly indicating the non-commutativity between the CM position coordinates with $\theta=\frac{m_{+}^2-m_{-}^2}{(m_{+}+m_{-})^2}il_B^2\epsilon_{ij}$ being the effective non-commutativity parameter. However, it is straightforward to check that the other two commutation brackets remain preserved. 
\begin{equation}[\hat{P}_{i},\hat{P}_{j}]=0;~~ [\hat{R}_{i},\hat{P}_{j}]=i\hbar \delta_{ij}
\end{equation}
It may be noted that the order of magnitude of the non-commutativity between the CM position coordinates is much lesser compared to that of the position coordinates of the individual constituent particles. This is simply because $l_B^2$ itself is very small due to the strong magnetic field limit, the presence of the additional mass factor reduces the whole effective non-commutativity parameter $\theta$ to a much smaller value.

Now, let us introduce the relative coordinate system:
\begin{align}
\scalebox{0.92}{$\hat{r}_{i}=\hat{y}_{i}-\hat{x}_{i};~~\hat{\Tilde{P}}_{i}=\frac{m_{+}}{m_{+}+m_{-}}\hat{P}_{y_{i}}-\frac{m_{-}}{m_{+}+m_{-}}\hat{P}_{x_{i}};~~i=1,2$}
\end{align}
The commutation relations satisfied by the relative coordinates are given by
\begin{align}
\scalebox{0.85}{$[\hat{r}_{i},\hat{r}_{j}]=0;~~[\hat{\Tilde{P}}_{i},\hat{\Tilde{P}}_{j}]=\frac{m_{-}^2-m_{+}^2}{(m_{+}+m_{-})^2}i\frac{\hbar^2}{l_{B}^2}\epsilon_{ij};~~[\hat{r}_{i},\hat{\Tilde{P}}_{j}]=i\hbar\delta_{ij};~~i,j=1,2 $}
\end{align}
It is  evident that the relative position coordinates commute as we have considered two oppositely charged particles on a non-commutative space (it has been shown earlier \cite{pmho}, that the non-commutativity of a charged particle differs from its antiparticle and also from any other particle of opposite charge by the sign). On the other hand, the coordinates of relative momenta give rise to a nontrivial commutation algebra with a reduced order of magnitude from that of the individual constituent particle's momentum coordinates.

It may be further noted that the position coordinates of the centre of mass and the position coordinates of the relative motion are not independent, rather they obey the relation given by 
\begin{equation}
[\hat{R}_{i},\hat{r}_{j}]=-i{l_{B}^2}\epsilon_{ij};~~i,j=1,2
\end{equation}
So, clearly, there is a connection between the motion of the centre of mass and the relative motion of the composite system in the non-commutative space. This helps us to reduce the two-body problem completely to a one-body problem for the internal motion in non-commutative space using the CM coordinates of the composite system where the information of the negatively charged particle is solely hidden/encoded within the CM momenta giving rise to a standard commutative algebra.

\section{Appendix B}

Here we provide a derivation for the expression of the purity function. We begin with the  expression of the reduced density matrix of the equation (\ref{reduced}) and the expression of the coherent state $\ket{\alpha(k_2)}$ and definition of the Purity function from the equation(\ref{purityfn}),
\begin{equation}
    \text{P}(\alpha)=\sum_l \sum_s \bra{l}\hat{\rho}_{red}\ket{s}\bra{s}\hat{\rho}_{red}\ket{l} \nonumber
\end{equation}
\begin{align}
\label{ap1}
   \scalebox{0.87}{$ \bra{l}\hat{\rho}_{red}\ket{s}=\frac{\sigma}{\sqrt{\pi}}\int_{-\infty}^{+\infty}\bra{l}\ket{\alpha(k_2)}\bra{\alpha(k_2)}\ket{s}e^{-\sigma^2(k_2-k_0)^2}dk_2 $}
\end{align}
The coherent state  can be expressed  as $$\ket{\alpha(k_2)}=e^{-\frac{\alpha^2}{2}}e^{\alpha\hat{a}_1^{\dagger}} e^{-\alpha \hat{a}_1}\ket{0}=e^{-\frac{\alpha^2}{2}} e^{\alpha \hat{a}_1^{\dagger}}\ket{0}$$
\begin{equation}
    \bra{l}\ket{\alpha(k_2)}=\bra{l}e^{-\frac{\alpha^2}{2}}\sum_{n=0}^{\infty}\frac{\alpha^n}{\sqrt{n!}}\ket{n}=e^{-\frac{\alpha^2}{2}}\frac{\alpha^l}{\sqrt{l!}}
\end{equation}
Similarly, $\bra{\alpha(k_2)}\ket{s}=e^{-\frac{\alpha^2}{2}}\frac{\alpha^s}{\sqrt{s!}}$.
Plugging this into the equation (59), one gets
\begin{equation}
    \bra{l}\hat{\rho}_{red}\ket{s}=\frac{\sigma}{\sqrt{\pi}}\int_{-\infty}^{+\infty} e^{-\alpha^2}\frac{(\alpha)^{l+s}}{\sqrt{l!}\sqrt{s!}}e^{-\sigma^2(k_2-k_0)^2} dk_2
\end{equation}
Now substituting, $\alpha(k_2)=\beta k_2$,
where $\beta=\xi l_B$, we get-
\begin{align}
\scalebox{0.9}{$ \bra{l}\hat{\rho}_{red}\ket{s}=\frac{\sigma}{\sqrt{\pi}}\frac{\beta^{l+s}}{\sqrt{l!}\sqrt{s!}} e^{-\sigma^2 k_0^2} \int_{-\infty}^{+\infty} e^{-(\beta^2+\sigma^2)k_2^2+2\sigma^2k_0k_2} k_2^{l+s}dk_2 $}
\end{align}
\begin{align}
  \scalebox{0.9}{$ =\frac{\sigma}{\sqrt{\pi}}
    \frac{\beta^{l+s}}{\sqrt{l!}\sqrt{s!}} e^{-\sigma^2 k_0^2} 
    \frac{1}{(2\sigma^2)^{l+s}}\frac{\partial^{l+s}}{\partial k_0^{l+s}}\{\int_{-\infty}^{+\infty} e^{-(\beta^2+\sigma^2)k_2^2+2\sigma^2k_2k_0} dk_2\}$} 
\end{align}
\begin{equation}
\label{ap2}
    =\frac{\sigma}{\sqrt{\pi}}
    \frac{\beta^{l+s}}{\sqrt{l!}\sqrt{s!}} e^{-\sigma^2 k_0^2} 
    \frac{1}{(2\sigma^2)^{l+s}}\frac{\partial^{l+s}}{\partial k_0^{l+s}}\{
    \sqrt{\frac{\pi}{\beta^2+\sigma^2}}e^{\frac{\sigma^4k_0^2}{\beta^2+\sigma^2}} \}
\end{equation}
Using the above expressions in the purity function, we  get,
\begin{align}
    \scalebox{0.85}{$\text{P}(\alpha)= \frac{\sigma^2}{\beta^2+\sigma^2} e^{(-2\sigma^2 k_0^2)} 
  \sum_l \sum_s  \frac{1}{l! s!}(\frac{\beta^2}{4\sigma^4})^{l+s}
    [e^{\frac{\sigma^4k_0^2}{\beta^2+\sigma^2}} \overset{\leftarrow}{\frac{\partial^{l+s}}{\partial k_0^{l+s}}} \overset{\rightarrow}{\frac{\partial^{l+s}}{\partial k_0^{l+s}}} e^{\frac{\sigma^4k_0^2}{\beta^2+\sigma^2}}]$}
\end{align}
Performing the summations, we are led to 
\begin{equation}
    \text{P}(\alpha(k_2))=\frac{\sigma^2}{\beta^2+\sigma^2} e^{(-2\sigma^2 k_0^2)} 
    [e^{\frac{\sigma^4k_0^2}{\beta^2+\sigma^2}} (e^{\frac{\beta^2}{2\sigma^4}\overset{\leftarrow}{\frac{\partial}{\partial k_0}}
    \overset{\rightarrow}{\frac{\partial}{\partial k_0}}})
    e^{\frac{\sigma^4k_0^2}{\beta^2+\sigma^2}}]
\end{equation}
Now, replacing $\xi_{0}=\frac{\xi l_{B}}{\sigma}$,  we arrive at-
\begin{equation}
\label{57}
    \text{P}(\xi_{0};l_{B})=(\frac{1}{1+\xi^{2}_{0}})e^{(-2\sigma^2 k_0^2)} [e^{(\frac{\sigma^2k_0^2}{1+\xi_{0}^{2}})} (e^{\frac{\xi^{2}_{0}}{2\sigma^2}\overset{\leftarrow}{\frac{\partial}{\partial k_0}} \overset{\rightarrow}{\frac{\partial}{\partial k_0}}})  e^{(\frac{\sigma^2k_0^2}{1+\xi_{0}^{2}})}]
\end{equation}
Subsequently, we express $[e^{\frac{\sigma^2k_0^2}{1+\xi_0^2}} (e^{\frac{\xi_0^2}{2\sigma^2} \overset{\leftarrow}{\frac{\partial}{\partial k_0}} \overset{\rightarrow}{\frac{\partial}{\partial k_0}}}) e^{\frac{\sigma^2k_0^2}{1+\xi_0^2}}]$ in a more concise form, where $\overset{\rightarrow}{\frac{\partial}{\partial k_0}}f=\frac{\partial f}{\partial k_{0}}$ and $f\overset{\leftarrow}{\frac{\partial}{\partial k_0}}=\frac{\partial f}{\partial k_{0}}$. For that, let us consider the following integral:
\begin{equation}
\int_{-\infty}^{+\infty} e^{-bs^2+2sk_0}ds=e^\frac{k_0^2}{b}\int_{-\infty}^{+\infty}e^{-b(s+\frac{k_0}{b})^2} ds=e^\frac{k_0^2}{b}\sqrt{\frac{\pi}{b}}
\end{equation}
From the expression of $e^{\frac{\sigma^2k_0^2}{1+\xi_0^2}}$, it follows that-
\begin{equation}
e^{\frac{\sigma^2k_0^2}{1+\xi_0^2}}=\sqrt{\frac{1+\xi_0^2}{\sigma^2\pi}}\int_{-\infty}^{+\infty} e^{-\frac{(1+\xi_0^2)}{\sigma^2}s^2+2sk_0}ds
\end{equation}
Therefore,
\begin{equation}
[e^{\frac{\sigma^2k_0^2}{1+\xi_0^2}} (e^{\frac{\xi_0^2}{2\sigma^2}\overset{\leftarrow}{\frac{\partial}{\partial k_0}} \overset{\rightarrow}{\frac{\partial}{\partial k_0}}})e^{\frac{\sigma^2k_0^2}{1+\xi_0^2}}]\nonumber
\end{equation}
\begin{align}
=\scalebox{0.9}{$ \frac{1+\xi_0^2}{\sigma^2\pi}\int_{-\infty}^{+\infty} e^{-\frac{(1+\xi_0^2)}{\sigma^2}s^2+2sk_0}ds(e^{\frac{\xi_0^2}{2\sigma^2}\overset{\leftarrow}{\frac{\partial}{\partial k_0}} \overset{\rightarrow}{\frac{\partial}{\partial k_0}}})\int_{-\infty}^{+\infty} e^{-\frac{(1+\xi_0^2)}{\sigma^2}s^{\prime 2}+2s^{\prime}k_0}ds^{\prime}$}
\nonumber
\end{align}
\begin{align}
\scalebox{0.9}{$ =\frac{1+\xi_0^2}{\sigma^2\pi}\int_{-\infty}^{+\infty} e^{-\frac{(1+\xi_0^2)}{\sigma^2}s^2+2sk_0}ds(e^{\frac{2\xi_0^2ss^{\prime}}{\sigma^2}})\int_{-\infty}^{+\infty} e^{-\frac{(1+\xi_0^2)}{\sigma^2}s^{\prime2}+2s^{\prime}k_0}ds^{\prime}$}
\nonumber
\end{align}
[where we have used the relation ${e^{a{\frac{\partial}{\partial k_0}}}{e^{bk_0}}}=e^{ab}e^{bk_0}$]
\begin{align}
\scalebox{0.95}{$ =\frac{1+\xi_0^2}{\sigma^2\pi}\int_{-\infty}^{+\infty} e^{-\frac{(1+\xi_0^2)}{\sigma^2}s^2+2sk_0}ds\int_{-\infty}^{+\infty} e^{-\frac{(1+\xi_0^2)}{\sigma^2}s^{\prime2}+2(k_0+\frac{\xi_0^2s}{\sigma^2})s^{\prime}}ds^{\prime}$}
\nonumber
\end{align}
\begin{equation}
=\sqrt{\frac{1+\xi_0^2}{\sigma^2\pi}}\int_{-\infty}^{+\infty} e^{-\frac{(1+\xi_0^2)}{\sigma^2}s^2+2sk_0}e^{\frac{(\sigma^2k_0+\xi_0^2s)^2}{\sigma^2(1+\xi_0^2)}}ds
\nonumber
\end{equation}
\begin{equation}
=\sqrt{\frac{1+\xi_0^2}{\sigma^2\pi}}e^{\frac{\sigma^2k_0^2}{1+\xi_0^2}}\int_{-\infty}^{+\infty} e^{-\frac{(1+2\xi_0^2)}{\sigma^2(1+\xi_0^2)}s^2+2k_0\frac{(1+2\xi_0^2)}{(1+\xi_0^2)}s}ds 
\nonumber
\end{equation}
After performing some suitable steps, we get the final simplified form as
\begin{equation}
[e^{\frac{\sigma^2k_0^2}{1+\xi_0^2}} (e^{\frac{\xi_0^2}{2\sigma^2}\overset{\leftarrow}{\frac{\partial}{\partial k_0}} \overset{\rightarrow}{\frac{\partial}{\partial k_0}}})
    e^{\frac{\sigma^2k_0^2}{1+\xi_0^2}}]=\frac{1+\xi_0^2}{\sqrt{1+2\xi_0^2}}e^{2\sigma^2k_0^2}   
\end{equation}
Now after plugging the above result (70) in equation (67), the expression of the Purity function reduces to
\begin{equation}
 \text{P}(\xi_{0}; l_{B})=\frac{1}{\sqrt{{1+2\xi^{2}_{0}}}}  
\end{equation}
which is the same as  equation (50).

\section{Appendix C}

Here we discuss how cat state-like behavior will be affected by generalizing the notion of impurity potential. To see this, we consider the system Hamiltonian in the presence of an impurity potential (a 2D oscillatory potential) attached to a positively charged particle:

\begin{equation}
\hat{H}_{NC}=\frac{\hat{P}^{2}_{1}}{2m_{B}}+\frac{\hat{P}^{2}_{2}}{2m_{B}}+V(\hat{x}_{1},\hat{x}_{2}),
\label{mhj}
\end{equation}
where \( V(\hat{x}_{1},\hat{x}_{2})=\frac{1}{2}K(\hat{x}^{2}_{1}+\hat{x}^{2}_{2}) \) and \( m_{B}=\frac{e^2B^2}{c^2K_{0}} \).

It may be noted that \( \hat{H}_{NC}=\hat{U}\hat{H}_{CM}\hat{U}^{\dagger} \), where

\begin{equation}
\hat{H}_{CM}=\frac{\hat{P}^{2}_{1}}{2m_{B}}+\frac{\hat{P}^{2}_{2}}{2m_{B}}+ \frac{1}{2}K\hat{R}^{2}_{1}+\frac{1}{2}K(\hat{R}_{2}+\frac{c}{eB}\hat{P}_{1})^{2}.
\label{hg}
\end{equation}

Here, $\hat{H}_{CM}$ indicates that the two modes are no longer independent, as the system Hamiltonian still retains its non-commutative effect through the explicit coupling between $\hat{R}_{2}$ and $\hat{P}_{1}$ when realized in terms of center-of-mass phase space variables. However, this system can be diagonalized by a suitable phase space transformation \cite{mu}. An important point is whether we can still achieve cat-like states. We can consider the specific states of \( \hat{H}_{CM} \), such as the ground state \( |\Psi_{g}\rangle_{CM} \), which satisfies $
\hat{H}_{CM}|\Psi_{g}\rangle = E_{g}|\Psi_{g}\rangle$. The ground state \( |\Psi_{g}\rangle \) can be written as a linear combination of the eigenstates of the Hamiltonian (\ref{mh}) as follows:

\begin{equation}
|\Psi_g\rangle_{CM} = \sum_n \int_{-\infty}^{\infty} dk_2 \langle n, k_2 | \Psi_g \rangle_{CM} | n, k_2 \rangle,
\end{equation}

where we have used the completeness relation for the eigenstates (\( |n, k_2\rangle = |n\rangle \otimes |k_2\rangle \)) of the Hamiltonian (\ref{mh}) for the system without potential along \( \hat{x}_{2} \).

When we transform the ground state \( |\Psi_g\rangle_{CM} \) using the unitary operator \( \hat{U} \), the operator \( \hat{U} \) introduces a superposition of displaced number states \cite{mul}. Specifically, this can be expressed as
$
|\alpha, n\rangle = \hat{U}|n, k\rangle = (e^{\alpha(\hat{a}_{1}^\dagger - \hat{a}_{1})}|n\rangle) \otimes |k_{2}\rangle,$ where $\alpha$ is a parameter associated with displacement (as discussed in (\ref{ncs}) for $n=0$). This transformation indicates that including a potential along the $x_2$ direction generally disrupts the coherent superposition of ``classical" (coherent) states. However, to realize cat-like states, it is essential to have a superposition of two diametrically opposite coherent states. Thus, to demonstrate the natural emergence of these cat-like states, we have considered the harmonic potential to depend solely on $\hat{x}_1$.

Fuethermore, if we consider the situation where the negatively charged particle is also attached to a harmonic trap potential along the \( y_1 \) direction, the system Hamiltonian can be expressed as:

\begin{equation}
\tilde{\hat{H}}_{NC}=\frac{\hat{P}^{2}_{1}}{2m_{B}}+\frac{\hat{P}^{2}_{2}}{2m_{B}}+V_{1}(\hat{x}_{1})+V_{2}(\hat{y}_{1}),
\end{equation}
where \( V_1(\hat{x}_1) \) and \( V_2(\hat{y}_1) \) are oscillatory potentials. To focus on the basic situation, we do not write them explicitly.

Using the fact of (\ref{jn}), we can eliminate the degrees of freedom of the second charged particle in terms of the phase space variables of the first charged particle as 
\(\hat{y}_{1}=\hat{x}_{1}+\frac{c}{eB}\hat{P}_{2}\), and the above Hamiltonian can be rewritten as:

\begin{equation}
\tilde{\hat{H}}_{NC} = \hat{U} \tilde{\hat{H}}_{CM}\hat{U}^{\dagger}
\end{equation}

with

\begin{equation}
 \tilde{\hat{H}}_{CM}=\frac{\hat{P}^{2}_{1}}{2m_{B}}+\frac{\hat{P}^{2}_{2}}{2m_{B}}+V_{1}(\hat{R}_{1})+ V_{2}\left(\hat{R}_{1}+\frac{c}{eB}\hat{P}_{2}\right) 
 \label{KL}
\end{equation}

The structure of Hamiltonian (\ref{KL}) clearly indicates that the introduction of a potential associated with the second charged particle along the \( \hat{y}_{1} \) direction effectively induces a coupling between \( \hat{P}_{2} \) and \( \hat{R}_{1} \). This is quite similar to the case of Hamiltonian (\ref{hg}). Following the same logic applied to the case for the positively charged particle  also trapped by the potential along the \( \hat{x}_{2} \) direction, we can immediately conclude that the feature of the cat-like state will not be exhibited if we allow the potential for the second negatively charged particle along \( \hat{y}_{1} \) along with the potential attached to the first positively charged particle along \( \hat{x}_{1} \). 

Therefore, for simplicity and clarity in our analysis focused on the positively charged particle, we restrict the harmonic potential \( V \) to depend solely on \( x_1 \).


\begin{thebibliography}{0}
\bibitem{Zuk}
W. H. Zurek, ``Decoherence and the Transition from Quantum to Classical," Phys. Today 36-44 ~1991, Quantum Theory of Measurement, edited by J.
A. Wheeler and W. H. Zurek ~Princeton U.P. Princeton, 1983!, pp. 152-167.

\bibitem{pn2}
 S. Haroche and J. M. Raimond, in Cavity Quantum
Electrodynamics, edited by P. Berman (Academic Press,
New York, 1994), p. 123.
\bibitem{par}
E. Schr\"{o}dinger, Naturwissenschaften 23, 807, 823, 844(1935)

\bibitem{2}

 A. J. Leggett, ``Schr\"{o}dinger's Cat and Her Laboratory Cousins," Contemp. Phys. 25, 583-598 ~1984
 \bibitem{bose}
C. C. Gerry and P. L. Knight, Quantum superpositions and Schr\"{o}dinger cat states in quantum optics, Am. J. Phys. 65,
964 (1997)

\bibitem{pnr}
R. Penrose, On gravity's role in quantum state reduction, General Relativity and Gravitation 28, 581 (1996).
\bibitem{pr}
A. Vinante, R. Mezzena, P. Falferi, M. Carlesso, and A. Bassi,
Improved Noninterferometric Test of Collapse Models Using
Ultracold Cantilevers, Physical Review Letters 119, 110401
(2017).
\bibitem{hr}
B. Helou, B. J. J. Slagmolen, D. E. McClelland, and Y. Chen,
LISA pathfinder appreciably constrains collapse models, Physical Review D 95, 084054 (2017).

\bibitem{hr1}
Angelo Bassi, Kinjalk Lochan, Seema Satin, Tejinder P. Singh, and Hendrik Ulbricht, ``Models of wave-function collapse, underlying theories, and experimental tests", Rev. Mod. Phys. 85, 471

\bibitem{Rov}
 H. J. Kimble, M. Dagenais, and L. Mandel, Phys. Rev. Lett.
39, 691 (1977).
\bibitem{mar}
R. Short and L. Mandel, Phys. Rev. Lett. 51, 384 (1983). 
\bibitem{mtw} 
 R. E. Slusher, L. W. Hollberg, B. Yurke, J. C. Mertz, and
J. F. Valley, Phys. Rev. Lett. 55, 2409 (1985).

\bibitem{qinform}
M.A. Nielsen and I.L. Chuang, Quantum Computation and Quantum Information,
Cambridge University Press, Cambridge, U.K. (2009).
\bibitem{qinf}
D. Bouwmeester, A. Ekert and A. Zeilinger, The Physics of Quantum Information
(Springer, Berlin, 2000).
\bibitem{qinfor}
A. Gilchrist, Kae Nemoto, W.J. Munro, T.C. Ralph, S. Glancy, Samuel. L. Braunstein and G.J. Milburn, Schr{\"o}dinger cats and their power for quantum information processing, J. Opt. B: Quantum Semiclass. Opt. 6, S828 (2004).

\bibitem{qmeter}
K. Gietka, Squeezing by critical speeding up: Applications
in quantum metrology, Phys. Rev. A 105, 042620
(2022).
\bibitem{qteleport}
Hung Do, Robert Malaney Jonathan Green, Teleportation of a Schr{\"o}dinger's-Cat State via Satellite-based Quantum Communications, arXiv:1911.04613v1 [quant-ph] 11 Nov 2019.




\bibitem{qerror1}
David S. Schlegel, Fabrizio Minganti, Vincenzo Savona, Quantum error correction using squeezed Schr{\"o}dinger cat states, Phys.Rev.A 106 (2022) 2, 022431
\bibitem{qerror2}
Jacob Hastrup and Ulrik Lund Andersen, All-optical cat-code quantum error correction, Phys. Rev. Research 4, 043065(2022) 

\bibitem{tw}
 W. Schleich,
J. P. Dowling, R.J. Horowicz, and S. Varro, in New Frontiers
in Quantum Optics and Quantum Electrodynamics, edited by
A. Barut (Plenum, New York, 1990)

\bibitem{legg}
The notion of interference between macroscopically distinguishable states has been promoted most prominently by A.
Leggett, in Proceedings of the Internal Symposium on Foundations of Quantum Mechanics in the Light of New Technology, edited by S. Kamefuchi (Physical Society of Japan, Tokyo,1983).

\bibitem{egg}
 G.J. Milburn and D. F. Walls, Phys.Rev. A 3S, 1087 (1988)
 
 \bibitem{Sb}
E. C. G. Sudarshan, Equivalence of semiclassical and quantum mechanical descriptions of statistical light beams, Phys. Rev. Lett. 10, 277 (1963).

\bibitem{kaku}
M. Cosacchi, J. Wiercinski, T. Seidelmann, M. Cygorek, A.
Vagov, D. E. Reiter, and V. M. Axt, On-demand generation of
higher-order Fock states in quantum-dot-cavity systems, Phys.
Rev. Research 2, 033489 (2020).

\bibitem{kau}
C. Navau, S. Minniberger, M. Trupke, and A. Sanchez, Levitation of superconducting microrings for quantum magnetomechanics, Physical Review B 103, 174436 (2021).


\bibitem{goenner}
 B. Li, W. Qin, Y. F. Jiao, C. L. Zhai, X. W. Xu, L. M.
Kuang, H. Jing, Optomechanical Schr\"{o}dinger cat states
in a cavity Bose-Einstein condensate, Fundamental Research, 3, 15 (2023).

\bibitem{rbm}
J. Foo, R. B. Mann, and M. Zych, Schr\"{o}dinger's cat for de Sitter spacetime, Classical and Quantum Gravity 38, 115010 (2021) 
\bibitem{wolf}
C. Marletto and V. Vedral, Gravitationally Induced Entanglement between Two Massive Particles is Sufficient Evidence
of Quantum Effects in Gravity, Physical Review Letters 119,
240042 (2017)
\bibitem{sch}
M. Christodoulou and C. Rovelli, On the possibility of laboratory evidence for quantum superposition of geometries, Physics
Letters, Section B: Nuclear, Elementary Particle and High Energy Physics 792, 64 (2019).
\bibitem{sch1}
 J.-Q. Liao, J.-F. Huang, and L. Tian, Generation of macroscopic
Schr\"{o}dinger-cat states in qubit-oscillator systems, Phys. Rev. A
93, 033853 (2016).
\bibitem{sch2}
F.-X. Sun, S.-S. Zheng, Y. Xiao, Q. Gong, Q. He, and K.
Xia, Remote Generation of Magnon Schr\"{o}dinger Cat State via
Magnon-Photon Entanglement, Phys. Rev. Lett. 127, 087203
(2021).

\bibitem{ed}
R. J. Marshman, S. Bose, A. Geraci, and A. Mazumdar, ``Entanglement of Magnetically Levitated Massive Schr\"{o}dinger Cat States by Induced Dipole Interaction,"  arXiv:2304.14638.

\bibitem{ein}
S. Doplicher, K. Fredenhagen and J. Roberts, {\em Commun. Math.Phys. 172 (1995) 187};
{\em Phys. Lett. B331 (1994) 39}

\bibitem{in}
H. Snyder, Quantized space-time, Phys. Rev. 71 (1) (1947) 38-41
\bibitem{go}
  A. Connes, Noncommutative geometry and reality, J.
Math. Phys. 36, 6194 (1995); R. Szabo, Phys.Rep. 378 (2003) 207; E. Akofor, A. P. Balachandran and A. Joseph, arXiv:0803.4351 (hep-th).
\bibitem{pe}
 R. Banerjee, B. Chakraborty, S. Ghosh, P. Mukherjee and S. Samanta, ``Topics in Noncommutative Geometry Inspired Physics." , Found Phys 39, 1297 (2009).
 \bibitem{pek}
 M. Chaichian, P. P. Kulish, K. Nishijima, and A. Tureanu,
On a Lorentz-invariant interpretation of noncommutative
space-time and its implications on noncommutative QFT,
Phys. Lett. B 604, 98 (2004).
\bibitem{ek}
S. Girvin, cond-mat/9907002; N. Macris and S. Ouvry, J. Phys. A 35, 4477 (2002).
\bibitem{qhefluids}
 L. Susskind, The Quantum Hall fluid and noncommutative Chern-Simons theory,
2001, [arXiv:hep-th/0101029]



\bibitem{vpn}
R. Jackiw and V.P. Nair, Phys. Rev. D43 (1991) 1933.
\bibitem{Rabin}
N. Banerjee, R. Banerjee and S. Ghosh, 
``Relativistic theory of free anyon revisited" Phys. Rev. D54, 1719 (1996).
\bibitem{k}
H. Ishizuka, N. Nagaosa, ``Noncommutative quantum mechanics and skew scattering in
ferromagnetic metals", Phys. Rev. B 96, 165202 (2017)
\bibitem{fg1}
IB Pittaway and FG Scholtz, ``Quantum interference
on the non-commutative plane and the quantum-to-classical transition", J.Phys.A 56 (2023) 16, 165303
\bibitem{fg2}
D Trinchero and  FG Scholtz, ``Pinhole interference in three-dimensional fuzzy space" Annals Phys. 450 (2023) 169224
\bibitem{kk}
G. Amelino-Camelia, L. Doplicher, S. Nam and Y.-S. Seo, Phys. Rev. D 67, 085008 (2003)
\bibitem{c}
I. Hinchliffe and N. Kersting, Int. J. Mod. Phys. A 19, 179 (2004)
\bibitem{d}
G. Amelino-Camelia, G. Mandanici and K. Yoshida, J. High Energy Phys. 01, 037 (2004)
\bibitem{qhe}
Z. Dong and T. Senthil, Non-commutative field theory and composite Fermi Liquids in some quantum Hall systems, Phys. Rev. B 102, 205126 (2020)
\bibitem{e}
S. Hellerman and M. Van Raamsdonk, Quantum Hall
physics = noncommutative field theory, JHEP 2001
(10), 039, arXiv:hep-th/0103179.


\bibitem{exci}

K. Cong, G. T. Noe II, and J. Kono, Excitons in
magnetic fields, in Encyclopedia of Modern Optics (Second Edition) (Elsevier, Oxford, 2018) pp. 63-81
\bibitem{R1}

J. Gea-Banacloche, ``Collapse and revival of the state vector in the Jaynes-Cummings model: An example of state preparation by a quantum apparatus", Phys. Rev. Lett. 65, 3385 (1990).
\bibitem{yueberly}
Ting Yu and J.H. Eberly, Finite-Time Disentanglement via Spontaneous Emission, 
Phys. Rev. Lett. 93, 140404 (2004).






\bibitem{R2}

Saha, P., Majumdar, A. S., Singh, S., Nayak, N., 2010, ``Collapse and revival of atomic entanglement in an intensity dependent Jaynes-Cummings interaction", Int. J. Quant. Information, 8, 1397-1409

\bibitem{Dunne}

G.V.Dunne, R.Jackiw, C.A.Trugenberger, Phys.Rev D41(1990) 661

\bibitem{bag}
D. Bigatti, L. Susskind, Magnetic fields, branes and noncommutative geometry, Phys. Rev. D 62 (2000) 066004

\bibitem{pn}

P. Nandi, S. Sahu, S. K. Pal , Nucl. Phys. B 971 (2021) 115511
\bibitem{exmol}
 J. Zhou, W.-Y. Shan, W. Yao, and D. Xiao, Phys. Rev. Lett. 115,
166803 (2015).
\bibitem{emg}
A. Chernikov, T. C. Berkelbach, H. M. Hill, A. Rigosi, Y. Li,
O. B. Aslan, D. R. Reichman, M. S. Hybertsen, and T. F. Heinz,
Phys. Rev. Lett. 113, 076802 (2014)
\bibitem{dipex}
Cong K, Noe G T II, Kono J. Excitons in Magnetic Fields. Oxford:
Elsevier, 2018, p.63-81
\bibitem{BD}

E. Cobanera, P. Kristel, and C. Morais Smith,
Phys. Rev. B 93, 245422 (2016).
\bibitem{fj}
L. D. Faddeev and R. Jackiw, ``Hamiltonian reduction of unconstrained and
constrained systems," Phys. Rev. Lett. 60, 1692-1694 (1988)

\bibitem{Rb}
R. Banerjee, H. J. Rothe and K. D. Rothe, Phys. Lett.B 462 (1999) 248-251, R. Banerjee, The commutativity principle and lagrangian symmetries, arXiv:hep-th/0001087

\bibitem{pnb2}
S. Biswas, P. Nandi, B. Chakraborty, Phys.Rev.A, 102 (2020) 2, 022231.

\bibitem{n2}
H.J. Rothe, K.D. Rothe, Classical and Quantum Dynamics of Constrained Hamiltonian Systems, World Scientific, Singapore, 2010.
\bibitem{hm2}
Bertolami, O.; Rosa, J.G.; de Aragao, C.M.L.; Castorina, P.; Zappala, D. Noncommutative gravitational quantum well. Phys. Rev. D 2005, 72, 025010.


\bibitem{pbom}
G. D. Barbosa and N. Pinto-Neto, ``Noncommutative quantum mechanics and Bohm's ontological interpretation", Phys. Rev. D 69, 065014 (2004).

\bibitem{pbom1}
G.D.Barbosa, J. High Energy Phys. 0305 (2003) 024
\bibitem{g}
P. Basu, B. Chakraborty and F. G. Scholtz, J. Phys. A 44 285204 (2011).
\bibitem{bom1}
W.M. Zhang, D. H. Feng, and R. Gilmore, ``Coherent states: Theory and some applications", Rev. Mod. Phys., vol. 62, pp. 867-927, Oct 1990.




\bibitem{scv}
 S. Mancini, D. Vitali, and P. Tombesi, Phys. Rev. Lett. 80, 688, 1998.

 \bibitem{sv}
 R.L. de Matos Filho and W. Vogel, ``Even and odd coherent states of
the motion of a trapped ion," Phys. Rev. Lett. 76, 608-611 (1996).

 \bibitem{gl}
C. Gerry and P. L. Knight, Introductory Quantum Optics (Cambridge University Press, Cambridge, 2004).

 \bibitem{v}
B. Yurke and D. Stoler, Generating quantum mechanical superpositions of macroscopically distinguishable states via amplitude dispersion, Phys. Rev. Lett. 57, 13 (1986).

 \bibitem{s}
A. Mecozzi and P. Tombesi, Phys. Rev. Lett. 58, 1055 (1987)

\bibitem{vn}
M. W. Noel and C. R. Stroud, Jr., in ``Coherence and
Quantum Optics VII," edited by J. Eberly, L. Mandel, and E. Wolf (Plenum, New York, to be published).


\bibitem{op}

J. Janszky, P. Domokos, and P. Adam, Coherent states on
a circle and quantum interference, Phys. Rev. A 48, 2213
(1993).
\bibitem{cohsup1}
R. Mirman, Analysis of the Experimental Meaning of Coherent Superposition and the Nonexistence of Superselection Rules, Phys. Rev. D 1, 3349 (Published: 15 June 1970).
\bibitem{cohsup2}
Mark R. Dowling, Stephen D. Barlett, Terry Rudolph, and Robert W. Spekkens, Observing a coherent superposition of an atom and a molecule, arXiv:quant-ph/0606128v2 11 Dec 2006.



\bibitem{kop}
Binayak Dutta Roy, ``Elements of Quantum Mechanics",  New Age Science (April 15, 2009)
\bibitem{rop1}
S. Bose, K. Jacobs, and P. L. Knight, Preparation of nonclassical
states in cavities with a moving mirror, Phys. Rev. A 56, 4175
(1997).
\bibitem{purity}
Gerardo Adesso, Alessio Serafini, and Fabrizio Illuminati, Entanglement, Purity, and Information Entropies in Continuous Variable Systems, arXiv:quant-ph/0506049v1 6 Jun 2005.

\bibitem{Dj6}
Brown, L.S.: Quantum motion in a Paul trap. Phys. Rev. Lett 66, 527 (1991).
\bibitem{Dj1}
Crefeld, C.E., Platero, G.: ac-driven localization in a two-electron quantum dot molecule. Phys. Rev. 
B 65, 113304 (2002).
\bibitem{Dj5}
Zeng, H.: Quantum-state control in optical lattices. Phys. Rev. A 57, 388 (1997).


\bibitem{Dj}
Thouless D J, Quantization of particle transport, Phys. Rev. B 27, 6083 (1983).



\bibitem{Dj3}
 Burmeister, G., Maschke, K.: Scattering by time-periodic potentials in one dimension and its infuence 
on electronic transport. Phys. Rev. B 57, 13050 (1998).


\bibitem{Dj4}
Li, W., Reichl, L.E.: Transport in strongly driven heterostructures and bound-state-induced dynamic 
resonances. Phys. Rev. B 62, 8269 (2000).



\bibitem{o2}A. Messiah, Quantum Mechanics (North-Holland, Amsterdam, 1962), Vol. 2



\bibitem{o3} P. Nandi, B. R. Majhi, N. Debnath, and S. Kala, “Quantum ballet by gravitational waves: Generating
entanglement’s dance of revival-collapse and memory within the quantum system,” Phys. Lett. B,
vol. 853, p. 138 706, 2024.






\bibitem{rop2}
S. Das and G. S. Agarwal, J. Phys. B 42, 141003 (2009).



\bibitem{exciband2}
Jagdeep Shah , ``Excitons in Semiconductor Nanostructures"  (Springer, 1999).

\bibitem{exciband3}
E. J. Sie, J. W. McIver, Y.-H. Lee, L. Fu, J. Kong, and N.
Gedik, Optical Stark effect in 2D semiconductors, Proc.
SPIE Int. Soc. Opt. Eng. 9835, 129 (2016).

\bibitem{exciband1}
Thomas Mueller and Ermin Malic, Exciton physics and device application of two-dimensional transition metal dichalcogenide semiconductors, npj 2D Materials and Applications (2018)2:29.
\bibitem{exc}
Lifshitz, E. M., Pitaevskii, L. P. (1980). Statistical Physics, Part 2 (2nd ed.). Pergamon Press.

\bibitem{exciband5}
Moskalenko, S. A. E., Moskalenko, S.  Snoke, D. Bose-Einstein Condensation
of Excitons and Biexcitons: and Coherent Nonlinear Optics with Excitons
(Cambridge Univ. Press, 2000).

\bibitem{er}
R. Jackiw, Nucl. Phys. B Proc. Suppl. 108 (2002), 30-36 doi:10.1016/S0920-
5632(02)01302-6 [arXiv:hep-th/0110057 [hep-th]]
\bibitem{r}
For a  review, see, S. Girvin, cond-mat/9907002.
\bibitem{r+}
J. S. Blakemore, J. Appl. Phys. 53, R123 (1982)
\bibitem{cr}
C. Cohen-Tannoudji, B. Diu, and F. Lalo¨e, Quantum Mechanics, Vol. 1 (Wiley and Hermann,
Paris 1977)
\bibitem{exp}
Kh. P. Gnatenko, Phys. Rev. D. 99, 026009 (2019);
\bibitem{ep}
 T.-C. Lu, Z. Zhang, S. Vijay, and T. H. Hsieh, PRX Quantum 4,
030318 (2023)

\bibitem{e4}
G. Bir and G. Pikus, Symmetry and Strain-induced Effects
in Semiconductors, A Halsted Press Book (Wiley, New York,
1974).







\bibitem{xio}
Xiao, D., J. Shi, and Q. Niu, 2005, ``Berry phase correction to electron density of states in solids,"  Phys. Rev. Lett. 95, 137204,
\bibitem{xio2}
D. Xiao, M.-C. Chang and Q. Niu, ``Berry phase effects on electronic properties", Rev. Mod. Phys. 82, 1959-2007 (2010).
\bibitem{Be1}
 W. Yao and
Q. Niu, ``Berry Phase Effect on the Exciton Transport and on the Exciton Bose-Einstein Condensate" Phys. Rev. Lett. 101, 106401 (2008)

\bibitem{Be2}
 J. Zhou, W.-Y. Shan, W. Yao, and D. Xiao, ``Berry Phase Modification to the Energy Spectrum of Excitons" Phys. Rev. Lett. 115,
166803 (2015).

\bibitem{Be3}
 A. A. Allocca, D. K. Efimkin, and V. M. Galitski, Fingerprints
of Berry phases in the bulk exciton spectrum of a topological
insulator, Phys. Rev. B 98, 045430 (2018).

\bibitem{sp}
A. Bhunia, M. K. Singh, M. A. Huwayz, M. Henini, and
S. Datta, Materials Today Electronics 4, 100039 (2023),
2107.13518
\bibitem{ps}
Bhunia, A., Singh, M. K., Huwayz, M. A., Henini, M. and Datta, S. (2021). Tailoring Quantum Oscillations of Excitonic Schrodinger's Cats as Qubits [arXiv:2107.13518]

\bibitem{pmho}
P.-M. Ho, H.-C. Kao, Phys. Rev. Lett. 88 (2002) 151602.

\bibitem{mu}

B. Muthukumar and P. Mitra, “Noncommutative oscillatorsand the commutative limit,” Physical Review D, vol. 66, no. 2,article 027701, 2002
\bibitem{mul}
 F. A. M. de Oliveira, M. S. Kim, P. L. Knight, et al., Phys. Rev. A, 41, 2645 (1990)

\bibitem{gold}
Goldman H and Senthil T 2022 Phys. Rev. B 105 075130





\end{thebibliography}
\end{document}